\documentclass[pra,aps,10pt,superscriptaddress,twocolumn,floatfix]{revtex4-1}
\usepackage{graphicx,color,xcolor}
\usepackage[nice]{nicefrac}	
\usepackage{amsmath,amssymb,bm}
\usepackage{amsmath}
\usepackage{braket}
\usepackage[rightcaption]{sidecap}

\usepackage[plainpages=false,pdfpagelabels,colorlinks=true,linkcolor=red,urlcolor=blue,citecolor=blue,pdftitle={},pdfauthor={},pdfdisplaydoctitle=true,pdfduplex=DuplexFlipLongEdge]{hyperref}

\definecolor{darkred}{rgb}{0.90,0.2,0.2}
\definecolor{darkgreen}{rgb}{0,0.60,.2}
\definecolor{darkblue}{rgb}{0.1,0.3,1}
\definecolor{grey}{cmyk}{0,0,0,0.25}
\definecolor{orange}{cmyk}{0,0.6,0.8,0}

\usepackage{multibib}
\newcites{M}{Methods References}
\newcites{SI}{Supplemental Information References}

\begin{document}


%
%

\title{Generalized hydrodynamics in strongly interacting 1D Bose gases}

\author{Neel Malvania$^*$}
\affiliation{Department of Physics, The Pennsylvania State University, University Park, Pennsylvania 16802, USA}
\author{Yicheng Zhang$^*$}
\affiliation{Department of Physics, The Pennsylvania State University, University Park, Pennsylvania 16802, USA}
\author{Yuan Le}
\affiliation{Department of Physics, The Pennsylvania State University, University Park, Pennsylvania 16802, USA}
\author{Jerome Dubail}
\affiliation{Universit\'e  de  Lorraine,  CNRS,  LPCT,  F-54000  Nancy,  France\\
$^*$These authors contributed equally to this work}
\author{Marcos Rigol}
\affiliation{Department of Physics, The Pennsylvania State University, University Park, Pennsylvania 16802, USA}
\author{David S. Weiss}
\affiliation{Department of Physics, The Pennsylvania State University, University Park, Pennsylvania 16802, USA}
\begin{abstract}
\vspace*{0.5cm}
\end{abstract}
\maketitle

{\bf The dynamics of strongly interacting many-body quantum systems are notoriously complex and difficult to simulate. A new theory, generalized hydrodynamics (GHD), promises to efficiently accomplish such simulations for nearly-integrable systems~\cite{castro2016emergent, bertini2016transport}. It predicts the evolution of the distribution of rapidities, which are the momenta of the quasiparticles in integrable systems. GHD was recently tested experimentally for weakly interacting atoms~\cite{schemmer2019generalized},  but its applicability to strongly interacting systems has not been experimentally established. Here we test GHD with bundles of one-dimensional (1D) Bose gases by performing large trap quenches in both the strong and intermediate coupling regimes. We measure the evolving distribution of rapidities~\cite{wilson_malvania_20}, and find that theory and experiment agree well over dozens of trap oscillations, for average dimensionless coupling strengths that range from 0.3 to 9.3. By also measuring momentum distributions, we gain experimental access to the interaction energy and thus to how the quasiparticles themselves evolve. The accuracy of GHD demonstrated here confirms its wide applicability to the simulation of nearly-integrable quantum dynamical systems. Future experimental studies are needed to explore GHD in spin chains~\cite{jepsen2020spin}, as well as the crossover between GHD and regular hydrodynamics in the presence of stronger integrability breaking perturbations~\cite{vasseur2020a, Doyon2020}.}

In interacting many-body quantum systems, like electrons in a metal, the low-energy properties can often be described to a good approximation in terms of quasiparticles that travel almost freely and have a finite lifetime. In integrable systems, quasiparticles are not an approximation and they live forever. They allow one to efficiently compute the entire energy spectrum~\cite{bethe1931theorie, gaudin2014bethe} and to study quantum dynamics~\cite{CauxEssler_2013, Caux_2016}. When there is weak integrability breaking, the quasiparticle picture remains useful at all energies. In many situations, GHD~\cite{castro2016emergent, bertini2016transport} dramatically simplifies the study of dynamics by focusing on the evolution of the momenta of the quasiparticles, the rapidities. Specifically, GHD consists of coupled hydrodynamic equations that are based on two assumptions \cite{castro2016emergent, dubail2016more, bulchandani2017solvable, doyon2019lecture}. First, the system is viewed as a continuum of fluid cells, each of which is spatially homogeneous, can be described by an integrable model, and contains many particles. Second, the time variation is slow enough that each fluid cell is locally equilibrated to a generalized Gibbs ensemble (GGE) parameterized by its distribution of rapidities~\cite{rigol_dunjko_07, mossel2012generalized, caux2012constructing, vidmar_rigol_16}. Experimentally testing GHD is tantamount to determining how robust these two assumptions are in real quantum dynamical systems.

\begin{figure}[!t]
\includegraphics[width=1\columnwidth]{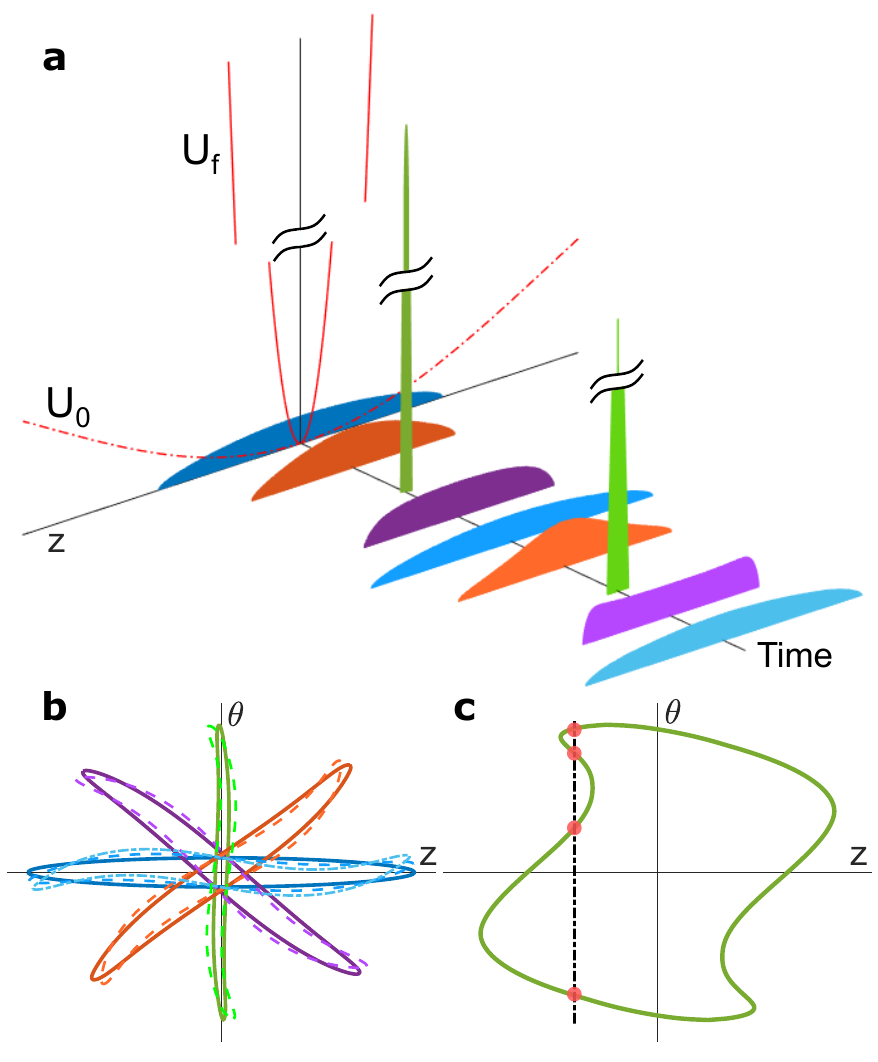}
\caption{\textbf{GHD theory after a 100-times trap quench.} \textbf{a}, Spatial evolution of the initially strongly coupled central 1D gas. At $t$=0, the trap depth, $U$, is increased by a factor of 100 (from the dashed to the solid red lines). The colored curves show the atomic density as a function of time as the cloud collapses and expands twice. \textbf{b}, The evolution in position($z$)-rapidity($\theta$) space. The colors correspond to the curves in \textbf{a}, with the solid lines showing the first cycle and the dotted lines the second cycle. \textbf{c}, A spatially magnified view of the half-period curve in \textbf{b}. The two segments of the dashed line inside the contour depict two Fermi seas, the second one of which (the small one) develops during the time evolution. Once a second Fermi sea forms, no existing theory other than GHD can model the evolution.}
\label{cartoon}
\end{figure}

In our experiments and GHD simulations, we suddenly ramp up the axial trapping potential around a bundle of ultracold 1D Bose gases and follow the evolution of the rapidity distribution as the gases successively collapse into the trap center and rebound (see Fig.~\ref{cartoon}a). Our study challenges the expectation that hydrodynamic descriptions only work for very large numbers of particles, since experiment and theory agree for weighted average numbers of atoms per 1D system as small as 11. The local equilibration condition is guaranteed to be initially satisfied since these are trap quenches; changes to the rapidity distribution only follow subsequent changes in the local density. However, we put the local equilibration condition to the test in our strongest trap quench. As the gases reach their maximum compression point we observe rapid (on the scale of the oscillation period) variations in the interaction energy that are associated with rapid changes in the nature of the quasiparticles.

In a 1D harmonic trap with no heating or loss, the dynamics of interacting bosons is approximately periodic; after one period the cloud returns to a configuration close to the initial one, slightly deformed by interactions during compression~\cite{caux2017hydrodynamics}. Assuming that the gas is initially in its ground state, analogous to a Fermi sea for quasiparticles~\cite{lieb1963exact}, it remains locally in a zero-temperature Fermi-sea state for some time. Such a situation could be described before the advent of GHD~\cite{de2016hydrodynamic, peotta2014quantum}. The Gaussian-shaped axial trap we use induces additional dephasing at the single-particle level, which enhances the differences between successive collapses. Eventually, GHD predicts~\cite{doyon2017large, caux2017hydrodynamics} the local formation of states with multiple Fermi seas~\cite{fokkema2014split}, illustrated in Fig.~\ref{cartoon}b and c, which show an example of phase space evolution in the first two cycles of one of our trap quenches. Such local states cannot be modeled with the methods that preceded GHD.  

To create a bundle of ultracold 1D gases from a $^{87}$Rb Bose-Einstein condensate confined by a 55~$\mu$m beam waist red-detuned crossed dipole trap, we slowly turn on a 2D blue-detuned optical lattice until it is 40$E_r$ deep, where $E_r = \hbar^2 k^2 / (2m)=2.55\times 10^{-30}J$ is the recoil energy defined by the lattice light wavevector, $k=2\pi/(772\,\text{nm})$~\cite{kinoshita_04}. We use two different initial conditions to study gases that are initially either intermediately or strongly 1D coupled, as characterized by the dimensionless parameter $\gamma=4.44/n_{{\rm 1D}}$, where $n_{{\rm 1D}}$ is the local 1D density in $\mu$m$^{-1}$ (see Methods). To start with an intermediate (strong) weighted average $\gamma$, i.e.~$\bar\gamma_0$ equals 1.4 (9.3), we use between 250,000 and 330,000 (83,000 and 119,000) atoms in a 9.4$E_r$ (0.56$E_r$) crossed dipole trap, so that the 2D distribution of 1D gases extends across a 17~$\mu$m (22~$\mu$m) radius. These radii are much smaller than the 2D lattice beam waists of 420~$\mu$m, so that the transverse confinement for each 1D gas is approximately the same. The axial trap depths, however, vary across the 1D gases, by up to 14$\%$ (27$\%$).

We measure rapidity distributions by first shutting off the axial trapping only and letting the atoms expand in 1D in a nearly flat axial potential until the momentum distributions have evolved into the rapidity distributions~\cite{wilson_malvania_20} (see Methods). Then we turn off the 2D lattice and measure the rapidity distributions via time of flight. Our previous dynamical fermionization measurement validated this momentum-to-rapidity mapping with parameters very close to our $\bar\gamma_0$=9.3 initial condition~\cite{wilson_malvania_20}. Because all the other initial cloud lengths are smaller, which allows for relatively more expansion in the flat potential, the mappings here are at least as good. We also measure momentum distributions by first suddenly turning off all the light traps. As the atoms rapidly expand transversely, atom interactions decrease rapidly and substantially~\cite{wilson_malvania_20}. The integrated axial spatial distributions after a time of flight reflect the 1D momentum distribution at the shut-off time. We adjust the times of flight for different measurements to maximize sensitivity, and simply rescale the spatial distributions to momentum.

\begin{figure}[!t]
\includegraphics[width=1\columnwidth]{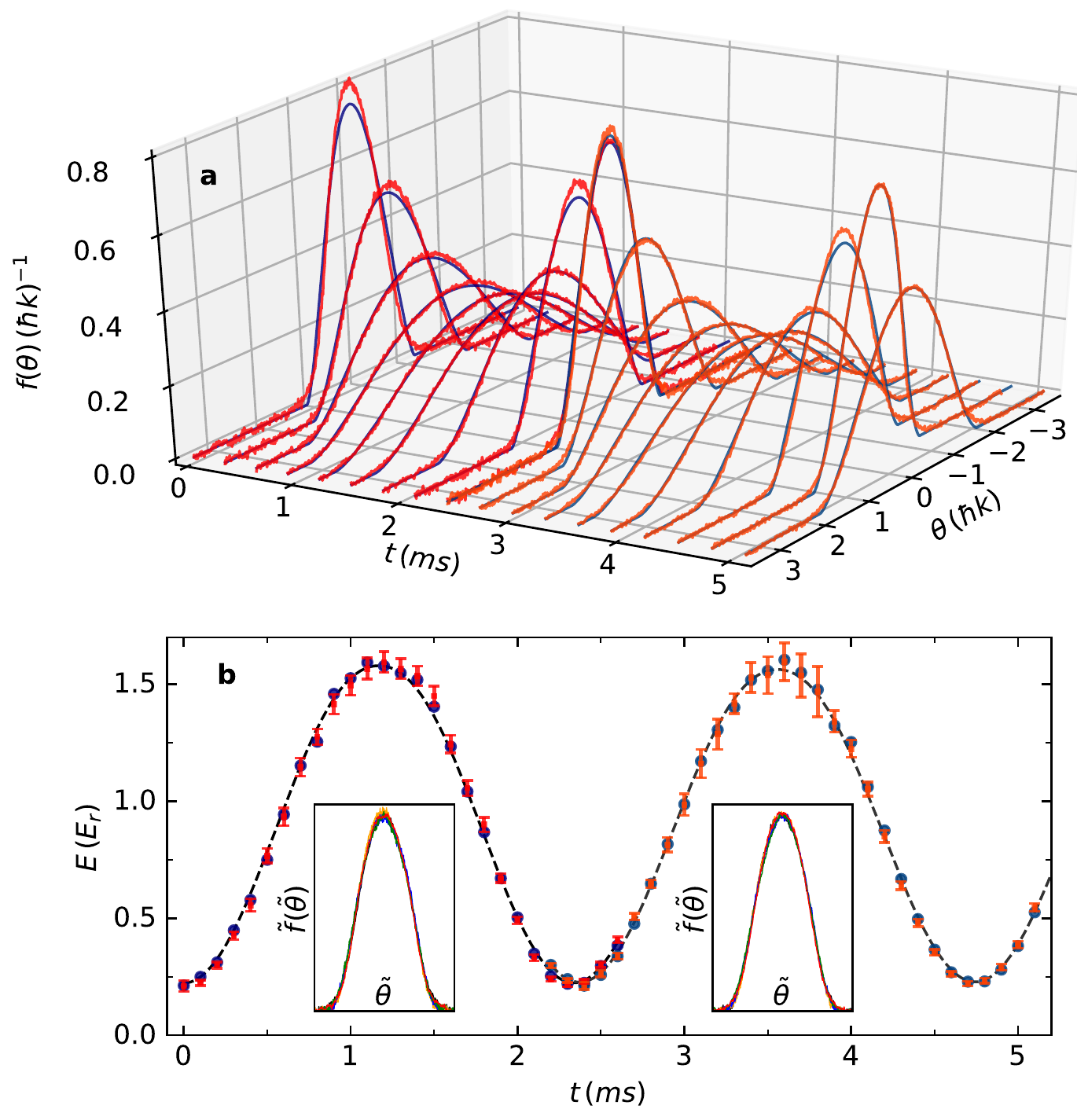}
\caption{\textbf{Intermediate coupling (10-times) trap quench.} \textbf{a}, Time evolution of the rapidity distribution after the trap is suddenly made ten times deeper for 1D gases with $\bar\gamma_0$=1.4. The red and orange curves show the experimental rapidity distributions over the course of the first two collapse cycles. The dark blue and light blue curves are the associated GHD theory, taking into account the measured atom number at each point (see Extended Data Fig.~1a and~b). The color change denotes a 2.7$\%$ change in the trap depth associated with the slow experimental drift (see Methods). \textbf{b}, Time evolution of the rapidity energy, $E$, after the quench. The red and orange squares are extracted from experimental distributions like those in \textbf{a} (see Methods). The dark and light blue circles are for the associated GHD theory. The dashed line is the GHD theory for the average atom number. The two insets show the rescaled experimental rapidity distributions ($\tilde f$ vs $\tilde \theta$, with arbitrary units) for points throughout the first and second cycle respectively (0, {$\pi$}/4, $\pi$/2, 3$\pi$/4, and $\pi$ phase points are black, orange, blue, green, and red, respectively). The curves are shape invariant.}
\label{10xquench}
\end{figure}

\newpage

\onecolumngrid

\begin{SCfigure}
\includegraphics[width=0.5\columnwidth]{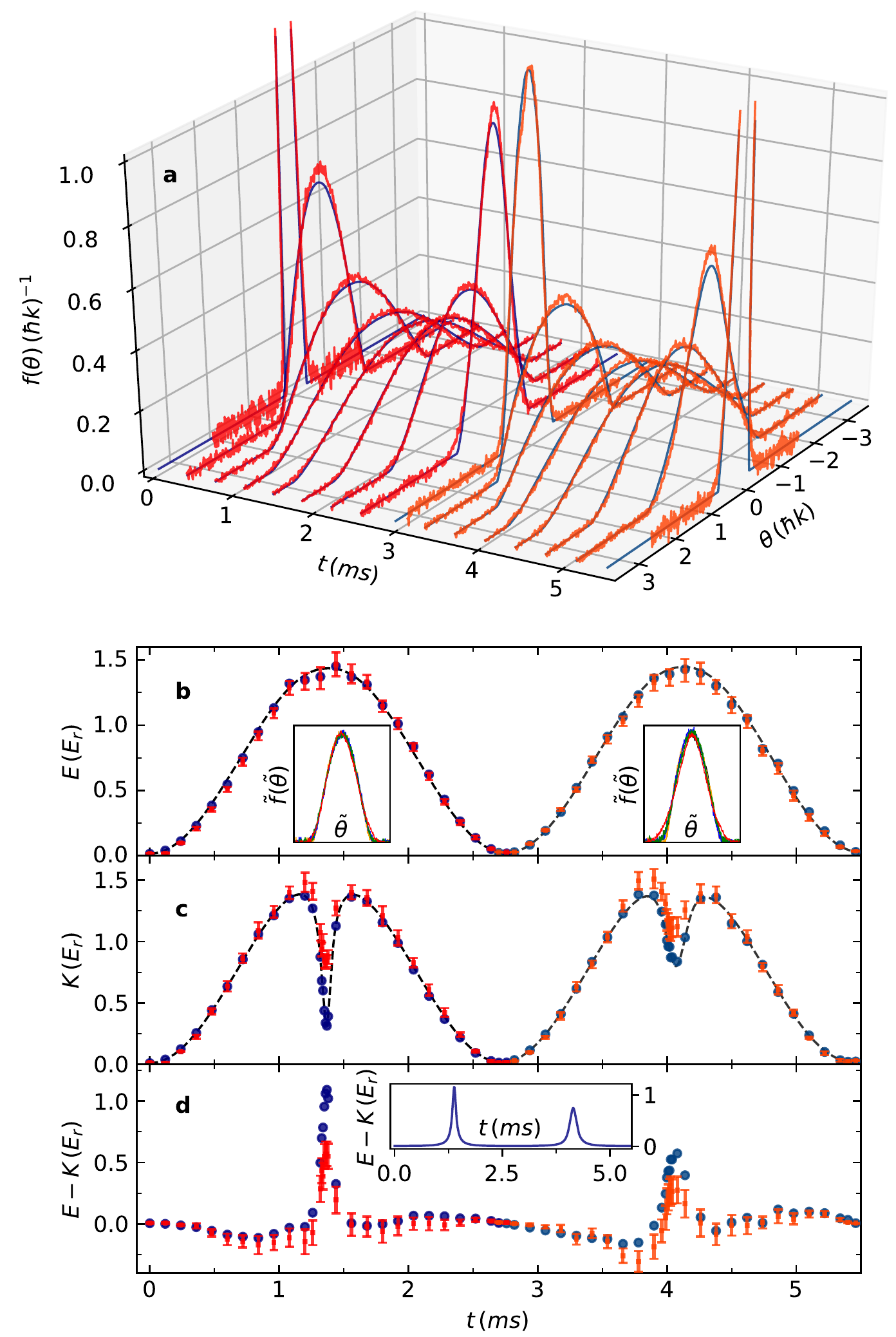}\hspace{0.1cm}
\caption{\textbf{Strong coupling (100-times) trap quench.} \textbf{a}, Time evolution of the rapidity distribution after the trap is suddenly made 100 times deeper for 1D gases with $\bar\gamma_0$=9.3. The red and orange curves show the experimental rapidity distributions over the course of the first two collapse cycles. The dark blue and light blue curves are the associated GHD theory, accounting for the measured atom number at each point (see Extended Data Fig.~1c and~d). The color change denotes a 1$\%$ change in the trap depth due to the slow experimental drift (see Methods). \textbf{b}, Time evolution of the rapidity energy, $E$, after the quench. The red and orange squares are extracted from experimental distributions like those in \textbf{a} (see Methods). The dark and light blue circles are for the associated GHD theory. The dashed line is the GHD theory using the average atom number. The two insets show the rescaled experimental rapidity distributions for points throughout the first and second cycle respectively (points near 0, $\pi$/4, $\pi$/2, 3$\pi$/4, and $\pi$ phases are shown in black, orange, blue, green, and red, respectively). By the second cycle the distributions are no longer self-similar. \textbf{c}, Time evolution of the kinetic energy, $K$, after the quench. The labeling is the same as for \textbf{b}, but the trap depths are slightly ($<4\%$) different (see Extended Data Fig.~1e and~f for the associated atom numbers). \textbf{d}, Time evolution of the interaction energy after the quench. The experimental and GHD theory points are obtained from \textbf{b} and \textbf{c} by subtracting $K$ from $E$ at each time. The inset shows GHD theory for a constant atom number and trap depth.\vspace{0.5cm}}
\label{100xquench}
\end{SCfigure}
\twocolumngrid

Figure~\ref{10xquench}a shows the evolution of the rapidity distribution starting from our intermediate coupling condition after a quench to a ten times deeper trap. Our quenches are small enough to ensure that two atoms never have enough energy to get transversely excited in a collision~\cite{riou_14}. Over the first two cycles, the shapes of all the distributions are self-similar (see Fig.~\ref{10xquench}b insets). Figure~\ref{10xquench}b shows the evolution of the integrated energy associated with the rapidities, which is the total energy less the trap potential energy. The squares are for the experiment, the dashed line shows the theory for an average number of atoms, and the circles show the theory using the measured number of atoms at each point (see Methods). After the quench, the calculated average cloud size drops from 14~$\mu$m to 3~$\mu$m, and $\bar\gamma$ drops from 1.4 to 0.3 (see Extended Data Fig.~2a--j). Figure~\ref{10xquench} clearly shows that GHD accurately describes these experiments, where the weighted average (maximum) number of atoms per 1D gas is 60 (140) and the nature of the quasiparticles changes gradually during the collapse. The onset of multiple Fermi seas for this setup occurs in the 3rd cycle. By the 11th cycle, we experimentally observe a loss of self-similarity that is consistent with our theoretical calculations. However, by that time a $\sim$20\% atom loss complicates the theory beyond the scope of this work~\cite{bouchoule2020effect} (see Extended Data Fig.~3).

Our initial strong coupling condition allows us to measure dozens of cycles without appreciable loss, and it also allows us to do a much larger trap quench, to a 100 times deeper trap. Figure~\ref{100xquench}a shows the rapidity evolution over the first two cycles. The shapes are no longer self-similar by the end of the first cycle (see the insets of Fig.~\ref{100xquench}b). The GHD theory agrees well with the experiment throughout. A second Fermi sea (see Fig.~\ref{cartoon}c) emerges during the first collapse; GHD is essential past that point. Extended Data Fig.~2k shows theoretical calculations of the evolution of cloud sizes; averaged over all tubes, the full width at half the central density decreases by a factor of 35, from 17.5~$\mu$m to 0.5~$\mu$m.

The squares, the dashed line, and the circles in Fig.~\ref{100xquench}b show the integrated rapidity energy as a function of time respectively for the experiment, the theory with the average atom number, and the theory with the measured atom numbers. The squares in Fig.~\ref{100xquench}c show the integrated kinetic energy as a function of time, determined from the measured momentum distributions. The momentum measurement near peak compression is somewhat compromised by the large interaction energy, some of which gets converted to kinetic energy early in the time-of-flight. There is no corresponding complication in the rapidity measurement. We extract the theoretical kinetic energy from GHD by using the Lieb-Liniger model in each GHD spatial cell to determine the interaction energy, integrating it, and subtracting it from the total integrated rapidity energy (see Methods). We adjust the axial trap depth in the theory to account for day to day experimental drifts ($<4\%$). The dashed line in Fig.~\ref{100xquench}c shows the result for the average atom number; the circles use the measured atom numbers. Figure~\ref{100xquench}d shows the difference between the rapidity and kinetic energies as a function of time, both experimentally and theoretically (with measured atom numbers), while the inset shows the theory for a fixed atom number and trap depth.

The data in Fig.~\ref{100xquench} probes the validity of GHD in two distinct ways. First, the weighted average (maximum) occupancy in the 1D gases is 11 (25), which is low enough to call the continuum approximation into question. We have compared GHD in the infinite $\gamma$ limit to exact theory (see Supplementary information) and find that if one filters out the spatial ripples that occur for small particle numbers, the GHD theory overlaps the exact one. Our experimental average over many 1D gases with different $N$ performs this filtering, and GHD reproduces it. Second, during the compression $\bar\gamma$ drops from 9.3 to 0.4 (see Extended Data Fig.~2m), with the final factor of 8 decrease occurring in the final 0.2~ms (right after the kinetic energy maximum). During this time the ratio of interaction energy to kinetic energy increases from 0.076 to 4.2, as illustrated by the first peak in Fig.~\ref{100xquench}d. The momentum distributions in this final stage of compression change shape from fermionic to bosonic, as was shown in Ref.~\cite{wilson_malvania_20} for a less dramatic quench. Our results validate the GHD description even in the face of rapid variations in the nature of the quasiparticles, suggesting that local GGE equilibration remains good throughout.

\begin{figure}[!t]
\includegraphics[width=1\columnwidth]{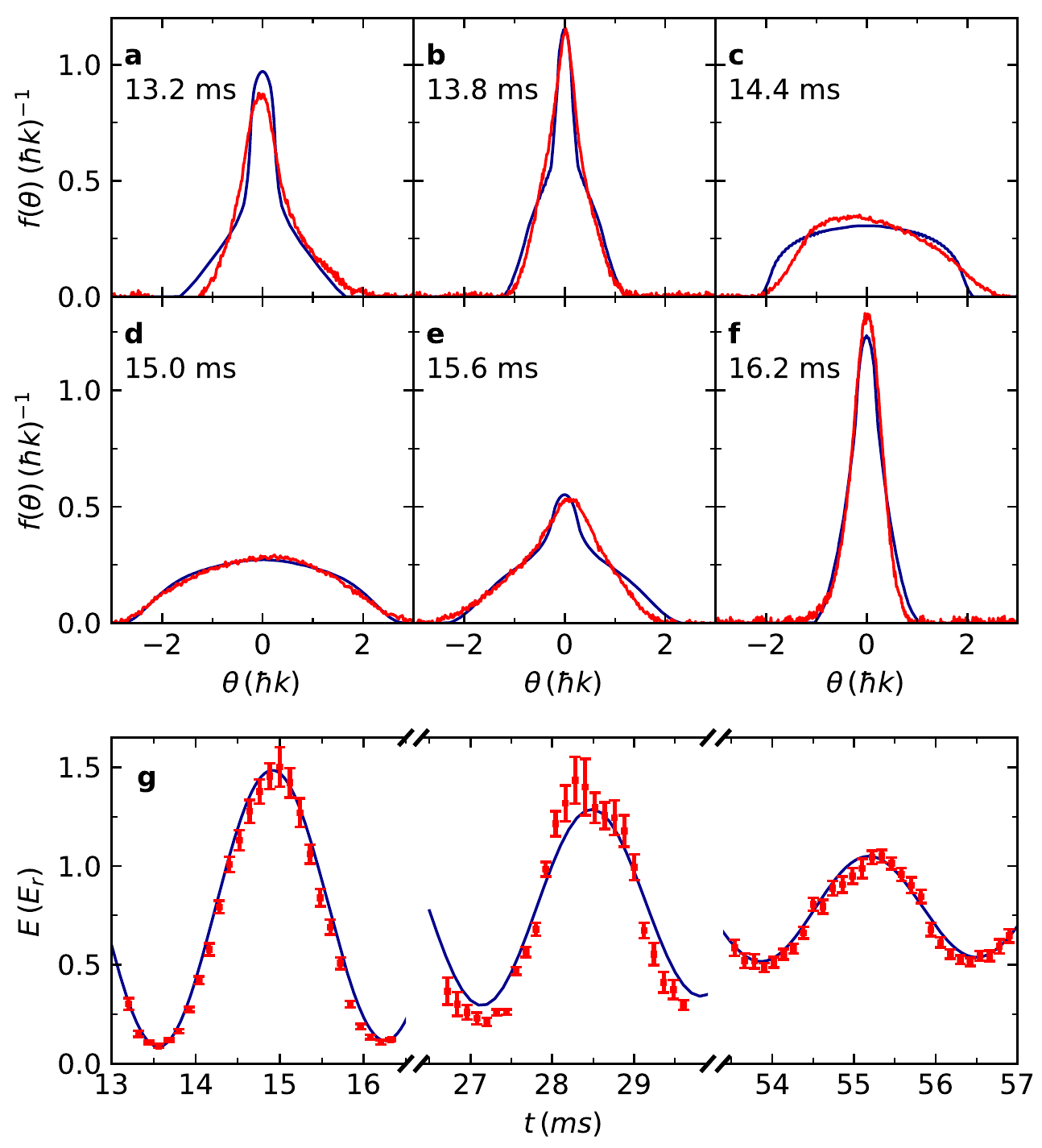}
\caption{\textbf{Strong coupling (100-times) trap quench at long times.} \textbf{a-f}, Experimental and GHD theory rapidity distributions during the sixth cycle after the 100 times quench. The red curves show the experiment, the blue curves the associated theory. \textbf{g}, The rapidity energy during the 6th, 11th and 21st oscillation after the quench. For all the theory curves in this figure we use the average number of atoms for each cycle, and the trap depth has been adjusted for each cycle so that the theory is in phase with the experiment.}
\label{sixthcycle}
\end{figure}

We next study rapidity distributions during the sixth oscillation cycle (see Fig.~\ref{sixthcycle}a--f). The theoretical distributions now change more noticeably during the cycle. Their distinctive shapes match the experimental curves reasonably well, except for slight experimental asymmetries due to drifts in the gravity-cancelling vertical magnetic field gradient, which lead to initial displacements of atoms from the light trap center by up to 100~nm, 0.006 of the full width of their distribution (see Methods). By the 11th cycle, finer features appear in the theory that are smeared out in the experiment (see Extended Data Fig.~4a--d), presumably on account of the initial atom cloud displacements and perhaps light trap asymmetry. In Fig.~\ref{sixthcycle}g, we show the integrated rapidity energies for cycles up to the 21st for the experiment and theory, with the period adjusted to account for small trap depth drifts (see Extended Data Fig.~4e--i for the corresponding rapidity distributions). The theory shows that there is $\lesssim$20\% loss of contrast in individual 1D gases, with the additional contrast loss coming from the inhomogeneity of axial trap depths. An extensive quantity, the rapidity energy is less sensitive to minor experimental imperfections. GHD is accurate enough to describe these experiments for the available measurement times.

We have shown that GHD accurately describes the dynamics of nearly integrable 1D Bose gases, with strong and intermediate coupling, after a trap quench. We experimentally challenged GHD's two underlying assumptions, the continuum and local GGE approximations, and we have done it for long evolution times. Natural next steps include performing similar tests on 1D systems that are farther from integrable, e.g., because of dipolar interactions~\cite{tang_kao_18} or dimensional cross-over~\cite{Schmiedmayer_trans2020}. Looking ahead, GHD and its extensions promise to become a standard tool in the description of strongly interacting 1D quantum dynamics close to integrable points. Such points describe a wide range of experiments involving 1D ultracold gases of bosons~\cite{cazalilla_citro_review_11} and fermions~\cite{guan2013fermi}, as demonstrated in experiments in continuum~\cite{kinoshita_wenger_06, pagano2014one, lev2020} and lattice~\cite{paredes_widera_04, fukuhara2013microscopic, Bloch2016, jepsen2020spin} systems.

\acknowledgements
Funding: Supported by NSF grants PHY-2012039 (D.S.W., N.M., and Y.L.), PHY-2012145 (Y.Z. and M.R.), and by U.S. Army Research Office grant W911NF-16-0031-P00005 (D.S.W., N.M., and Y.L.). The computations were carried out at the Institute for Computational and Data Sciences at Penn State. Author contributions: N.M. and Y.L. carried out the experiments; Y.Z. carried out the theoretical calculations; D.S.W. oversaw the experimental work, and J.D. and M.R. oversaw the theoretical work. All authors were involved in the analysis of the results, and all contributed to writing the paper.

\bibliographystyle{biblev1}
\bibliography{references}

%
%

\vspace{1cm}
\begin{center}
{\bf \large Methods}
\end{center}

\section{Initial state} \label{model}

{\bf Model.} The experimental system is modeled as a 2D array of 1D gases (dubbed ``tubes'' in what follows). Each tube ``$\ell$'' is described by the Lieb-Liniger model~\cite{lieb1963exact} in the presence of a confining potential $U_\ell(z)$,
\begin{equation}\label{H_lieb_liniger}
{\cal H}^\ell_\text{LL}=\sum_{j=1}^{N_\ell}\left[-\frac{\hbar^2}{2m}\frac{\partial^2}{\partial z^2_j}+U_\ell(z_j)\right]+g\sum_{1\leq j < l \leq N}\delta(z_j-z_l) \,,
\end{equation}
where $m$ is the mass of a $^{87}$Rb atom, $N_\ell$ is the number of atoms in tube $\ell$, and $g$ is the effective 1D contact interaction (repulsive in our case, so $g>0$)~\citeM{olshanii1998atomic}. $g$ depends on the depth of the 2D optical lattice $U_\text{2D}$, $g=-\hbar^2/(ma_\text{1D})$, where $a_\text{1D}=-a_\perp(a_\perp/a_s-C)/2$ is the 1D scattering length, $a_s$ is the (3D) $s$-wave scattering length, $a_\perp=\sqrt{2\hbar/(m\omega_\perp)}$ is the length of the transverse confinement ($\omega_\perp=\sqrt{2U_\text{2D}k^2/m}$), and $C= -\zeta(1/2) \simeq1.4603$~\citeM{olshanii1998atomic}. In the absence of the axial potential $U_\ell(z)$, the Hamiltonian above is exactly solvable via the Bethe ansatz \cite{lieb1963exact},~\citeM{yang1969thermodynamics}. In the ground state, observables depend only on the dimensionless coupling strength $\gamma=mg/(\hbar^2 n_{\rm 1D})$, where $n_{\rm 1D}$ is the 1D particle density.

In our experimental setup, the axial potential $U_\ell(z)$ for each tube varies as a function of its spatial location $(x_\ell,y_\ell)$ in the 2D array. We model it as the sum of the two Gaussian-shaped trapping beams propagating in the $x$ and $y$ directions, and the approximately harmonic anti-trapping potential caused by the blue detuned 2D lattice,
\begin{equation}\label{Trap_shape}
U_\ell(z)=U\left[1-\frac{1}{2}(e^{-\frac{x^2_\ell}{2W^2}}+e^{-\frac{y^2_\ell}{2W^2}})e^{-\frac{z^2}{2W^2}}\right]-\frac{1}{2} m \omega_\text{at}^2z^2\,,
\end{equation}
where $U$ ($W$) is the amplitude (width) of the Gaussian trapping beam, and $\omega_\text{at}$ is the frequency of the anti-trap. Initially, the atoms are assumed to be in the ground state of Eq.~(\ref{H_lieb_liniger}) in the presence of a potential with amplitude $U_{0}$ and width $W$. For the initial intermediate coupling condition, we use $U_{0}=19.39\,E_r$ and $W =55.2\,\mu$m, while for initial strong coupling condition, we use $U_0=1.195\,E_r$ and $W=55.2\,\mu$m. The 2D optical lattice depth after the state preparation is $40E_r$ for both quenches, which gives us $g=3.8\times10^{-37}$~Jm. The measured anti-trap frequency is $\omega_\text{at}=41.6$~s$^{-1}$. Since $\omega_\text{at}$ is much smaller than the Gaussian trapping beam frequency, we neglect it in our calculations.

The total number of atoms loaded into the 2D optical lattice fluctuates between consecutive measurements and slightly drifts in time. We fully account for these measured instabilities theoretically. For each average total atom number $N_\text{tot}(t)$, for measurements carried out at a time $t$ after the trap quench, we determine the number of atoms in each tube $N_\ell(t)$. For that, we assume that the loading of the optical lattice in the experiment is adiabatic (at all times the system is in the ground state), and that the 3D system decouples into a 2D array of tubes at a lattice depth $U^*_\text{2D}$ (this lattice depth sets $g^*$ at the time of the decoupling). Although different parts of the system will in fact decouple at different lattice depths, this approximation seems to work well. Then, using the local density approximation (LDA) and the exact Lieb-Liniger solution for the ground state of the homogeneous system, we determine $N_\ell(t)$. We round the atom number $N_\ell(t)$ in each tube to the closest integer. 

$U^*_\text{2D}$ is our only free parameter. We run GHD simulations for different values of $U^*_\text{2D}$ (taken to be integers times $E_r$, the recoil energy) and compare the rapidity energy results to the experimental ones. Specifically, we compute a least squares difference between the experiment and the theory at a set of discrete times about the maximal compression point in the first oscillation period of both quenches we study (see Extended Data Fig.~5a and~b). The minimum in each case is selected as the optimal $U^*_\text{2D}$. We check that the simulated atom distributions among tubes are consistent with the (less accurate) experimental measurements of the transverse distribution among tubes for the corresponding quenches, and that they are close to 2D projections of 3D Thomas-Fermi distributions (see Extended Data Fig.~5c and~d).

{\bf Centering the atomic cloud.} A well-controlled trap quench requires that the atomic cloud be precisely centered in the crossed dipole light trap. Otherwise, when the trap depth is suddenly increased, the gas collapses asymmetrically. Centering requires that the sum of the forces due to gravity, the axial magnetic field gradient, and the 2D optical lattice axial intensity gradient add to zero. We prepare a quantum degenerate gas in the spin polarized $F=1, m_F = 1$ state in a gravity-cancelling axial magnetic field gradient, which is generated, along with a magnetic bias field, by two coils in an asymmetric anti-Helmholtz configuration. To fine tune the levitation, we use a very small BEC of $2\times10^4$ atoms in a very shallow crossed dipole trap, $0.05 E_r$ deep. We measure the change in the position of the center of the atom cloud between two time-of-flight (TOF) measurements, at 20 ms and at 90 ms. By finely adjusting the current of one of the coils, we reduce the position drift of the $\sim60$ $\mu$m cloud to  less than 5 $\mu$m. This corresponds to a residual potential of less than $1.88\times10^{-28}$ J/m, which shifts the center of the atom cloud in the initial strong coupling condition by less than 100 nm. For the initial intermediate coupling condition, where the trap is an order of magnitude deeper, the center shift is less than 10 nm.

\section{Trap quench}

{\bf Experimental protocol.} For the quench, our goal is to create a rectangular pulse shape for the axial trap depth. The intensity of the dipole trap beams, which determines the depth of the axial trap, is controlled by electro-optic modulators (EOMs) that are powered by a high-voltage amplifier. The speed of trap depth changes is limited by both the slew rate of the supply and thermal effects in the EOMs. If the control voltage is simply pulsed on to the required set value, the light intensity increases to about $80\%$ of its steady-state value in 25 $\mu$s, followed by a slower climb to steady-state in about 70 $ms$. To keep the trap depth constant in the tens of milliseconds after the quench, we engineer a sequence of between one and three exponential voltage control ramps (depending on evolution time) that, when stitched together, counteract the EOM and high voltage amplifier transients. We end up with a nearly rectangular intensity pulse that is constant to within $2\%$ of the measured value over the quench duration. During data acquisition the steady-state depth can change by up to $2\%$ as a result of slow drifts in the laser power and in the photodiode measurement of the power. The latter drift can lead to day-to-day trap depth changes of up to $4\%$.

{\bf Modeling.} At time $t=0$, we assume the depth of the potential in each tube is suddenly changed from $U_0$ to $U_f$ (a trap quench), with $U_f>U_0$, and let the system evolve. For the simulation of the dynamics after the 10-times trap quench, $U_f=206.9\,E_r$. Because of power-dependent refractive effects in the electro-optic modulators, the beam waist, $W$, changes in this quench to 57.1 $\mu$m. For the 100-times trap quench, $U_f=125.7\,E_r$, and $W$ remains at 55.2 $\mu$m.

\section{Generalized hydrodynamics}\label{methodGHD}

To model the dynamics in each tube after the quench, we solve the following GHD equation~\cite{castro2016emergent, bertini2016transport},~\citeM{doyon2017note},
\begin{equation}\label{GHD_eq}
\partial_t n_\ell+v^\text{eff}_{\ell}\partial_z n_\ell= [\partial_z U_\ell(z)] \,  \partial_{\theta} n_\ell\,,
\end{equation}
which fixes the dynamics of the Fermi occupation ratio $n_\ell(\theta,z,t)\in [0,1]$ of the quasiparticles with rapidity $\theta$ in fluid cells at position $z$ and time $t$. The effective velocity $v^{\rm eff}$ is defined in Eq.~(\ref{effective_velocity}) below. (We note that an extension of GHD that includes diffusive corrections has been developed recently~\citeM{de2018hydrodynamic, de2019diffusion}. However, the diffusive term vanishes in our zero-entropy states, see Supplementary information.) In what follows, for as long as the discussion refers to a single tube, we remove the tube label $\ell$ to simplify the equations. We bring it back when needed to account for the presence of multiple tubes. 

The Fermi occupation ratio is defined as the ratio of the local density of quasiparticles $\rho(\theta,z,t)$ to the local density of states $\rho_{\rm s} (\theta,z, t)$: $n(\theta,z,t) = \rho(\theta,z,t)/ \rho_{\rm s} (\theta,z,t)$.  The density of states is $\rho_{\rm s} (\theta) = \frac{1}{2\pi \hbar} 1^{\rm dr} (\theta)$~\cite{lieb1963exact}, where the constant function $1(\theta) = 1$, and the dressing of any function $f(\theta)$ is defined by the integral equation
\begin{equation}\label{dressing}
f^\text{dr}(\theta)=f(\theta)+\int \frac{d\alpha}{2\pi}\varphi(\theta-\alpha)n(\alpha)f^\text{dr}(\alpha), 
\end{equation}
with
\begin{equation}
\varphi(\theta-\alpha)= \frac{2mg/\hbar}{(mg/\hbar)^2+(\theta-\alpha)^2} .
\end{equation}
The key ingredient of the GHD equation (\ref{GHD_eq}) is the formula for the effective velocity of the quasiparticles, $v^{\rm eff}$~\cite{bertini2016transport, castro2016emergent, doyon2019lecture}. It can be thought of as the group velocity for a quasiparticle with energy $\varepsilon = \theta^2/(2m)$ and momentum $p = \theta$, dressed by the particles' interactions,
\begin{equation}
\label{effective_velocity}
v^\text{eff}(\theta)= \frac{(d \varepsilon /d\theta)^{\rm dr}}{(d p /d\theta)^{\rm dr}} =  \frac{1}{m} \frac{{\rm id}^\text{dr}(\theta)}{1^\text{dr}(\theta)},
\end{equation}
with $\text{id}(\theta)=\theta$ and $1(\theta)=1$.

We start our simulation from the ground state of the Lieb-Liniger model [with a confining potential $U(z)$], which is a state with zero entropy~\citeM{yang1969thermodynamics}. Since evolution under the GHD equation (\ref{GHD_eq}) does not create entropy, the system remains in a zero-entropy state at all times. Consequently, at each position $z$ and time $t$, the occupation function $n(\theta,z,t)$ corresponds to a Fermi sea~\cite{lieb1963exact}, or to multiple Fermi seas~\cite{fokkema2014split}. It has the form
\begin{equation}
    \label{Fermi_occupation}
    n(\theta) = \left\{ \begin{array}{ccl}
        1 &{\rm if} & \theta \in [ \theta_1 , \theta_{2} ] \cup \dots \cup [\theta_{2q-1}, \theta_{2q} ] \\
        0 && {\rm otherwise},
    \end{array} \right.
\end{equation}
where the Fermi points $\theta_i$, as well as the number of components $q$ of the local multiple Fermi sea, depend on $z$ and $t$~\cite{doyon2017large}. Since the local state depends continuously on position, the set of points $(z, \theta_i (z) )$ in phase space at any given time $t$ is a closed curve $\Gamma_t$ (see Fig.~1b and Extended Data Fig.~6). The full GHD equation (\ref{GHD_eq}) can then be recast into a simpler equation for the closed curve $\Gamma_t$. Parametrizing the curve $\Gamma_t$ as $(z_t(s) , \theta_t (s))$ for $s \in [0,2\pi]$, Eq.~(\ref{GHD_eq}) becomes \cite{doyon2017large},~\citeM{ruggiero2020quantum}
\begin{equation}
    \label{Zero_Entropy_GHD}
    \frac{d}{dt} \left[ \begin{array}{c}
        z_t(s) \\
        \theta_t(s) 
    \end{array} \right] = \left[ \begin{array}{c}
        v^{\rm eff} (z_t(s) , \theta_t(s),t) \\
        -  \partial_z U (z_t(s))
    \end{array} \right] .
\end{equation}
It is this equation that we use in our numerical simulations. We use $500$ discrete points to approximate the curve $\Gamma_t$, and a time step of $dt = 0.1\, \mu$s. At each time step, we increment the position of the discrete point $(z_s, \theta_s)$ [$s \in \{0,\, \frac{2\pi}{500},\, 2\frac{2\pi}{500},\, \dots,499 \frac{2\pi}{500} \}$] as $(z_s, \theta_s) \rightarrow (z_s + v^{\rm eff} dt, \theta_s -  \partial_z U dt )$. For each discrete point, $v^{\rm eff}$ is evaluated by first searching for all other local Fermi points in order to know the local Fermi occupation ratio (\ref{Fermi_occupation}), and then by using Eqs.~(\ref{dressing})--(\ref{effective_velocity}).

Initially, for the ground state of the trapped Lieb-Liniger gas obtained using the LDA, there is only one Fermi surface [i.e., $q=1$ in Eq.~(\ref{Fermi_occupation})] at all positions in the system (see Supplementary information). After some time, the curve $\Gamma_t$ eventually folds onto itself, and a double Fermi sea (i.e., $q=2$) appears at some positions $z$ (see Fig.~1b and Extended Data Fig.~6). Points with $q=3,\,4,\, \dots$ eventually appear in a similar fashion. Before the appearance of such multiple Fermi seas, the conventional hydrodynamic equations of Galilean fluids, which make use of the exact equation of state provided by the solution of the Lieb-Liniger model~\cite{lieb1963exact}, can be used to simulate the dynamics of the gas~\cite{peotta2014quantum, de2016hydrodynamic}. In fact, Eq.~(\ref{Zero_Entropy_GHD}) is equivalent to the conventional hydrodynamics of Galilean fluids in that case~\cite{doyon2017large}. However, when a new Fermi sea is formed, the conventional hydrodynamics of Galilean fluids predicts an unphysical shock wave~\cite{peotta2014quantum}. When there is more than one Fermi sea, the full GHD equations are needed to simulate the dynamics.

The average (per particle) spatially integrated distribution of rapidities is given by
\begin{equation}
    f(\theta, t) \, = \, \frac{1}{N_{\rm tot}(t)}\sum_\ell \int dz \, \rho_\ell(\theta,z,t),
\end{equation}
where $\rho_\ell(\theta,z,t)$ is the local density of quasi-particles in the $\ell$th tube, and the sum runs over all tubes. We note that while $\rho_\ell(\theta,z,t)$ is expected to be a smooth function in all tubes, with no short-wavelength features, $f(\theta, t)$ can have them as a result of the sum over tubes (see, e.g., the small ripples in Fig.~4c). As shown in Extended Data Fig.~7, for the particular case in Fig.~4c, rapid changes in the (smooth) local density of quasi-particles $\rho_\ell(\theta,z,t)$ in individual tubes produce ripples in $f(\theta, t)$.

We also compute the average (per particle) kinetic energy ($E_K$), interaction energy ($E_I$), and total rapidity energy ($E$) by evaluating the following integrals (see Supplementary information) and sums over all tubes:
\begin{eqnarray}
E_K(t)&=&\frac{1}{N_{\rm tot}} \sum_\ell\int dzd\theta \rho_\ell \left[v^{\rm eff}_\ell-\frac{\theta}{2m}\right]\theta\,,\\
E_I(t)&=&\frac{1}{N_{\rm tot}} \sum_\ell\int dzd\theta \rho_\ell \left[\frac{\theta}{m} -v^{\rm eff}_\ell \right]\theta\,,\\
E(t)&=&\frac{1}{N_{\rm tot}} \sum_\ell\int dxd\theta \rho_\ell \frac{\theta^2}{2m} \,.
\end{eqnarray}

\section{Measurements}\label{align}

{\bf Rapidity distributions.} To experimentally measure the rapidity distribution at different times after the trap quench, we expand the cloud in a ``flat'' potential in 1D. We used this procedure in Ref.~\cite{wilson_malvania_20} to observe the dynamical fermionization of the momentum distribution of 1D bosonic gases in the Tonks-Girardeau regime~\citeM{rigol_muramatsu_05a, minguzzi_gangardt_05}, tantamount to observing the transformation of momentum distributions into rapidity distributions.

To create a flat potential, we leave on the red-detuned crossed dipole trap at low power to compensate for the anti-trap from the blue-detuned lattice~\cite{wilson_malvania_20}. To precisely align the two beams that make up the crossed dipole trap with the lattice anti-trap, we first turn on the 2D lattice around a BEC in the crossed dipole trap. We then ramp down the RF power to the acousto-optic tunable filter that acts as tunable beam-splitter for the dipole trap beams. With the axial confinement now provided by only one dipole trap beam (beam X) we suddenly turn it off. The atoms expand in the tubes and we image them 45 ms later. We adjust the alignment of beam X until the expansion is symmetric about the trap center. We then align the other dipole trap beam (beam Y) to beam X by alternately trapping non-degenerate atoms in a single beam (either X or Y) and adjusting beam Y until its vertical center matches beam X. In this way both the X and Y beams are aligned to the center of the 2D lattices to within 5 $\mu$m. 

{\bf Atom number measurement.} For each experimental image, we use a principle component analysis algorithm (PCA) to remove interference patterns from the background~\citeM{segal2010buriedinfo}. After PCA, there remains a background with low spatial frequency and low amplitude. We average 15 transversely-integrated images to obtain a 1D profile for each time-step. To remove the background, we perform a 4th-order polynomial fit to the part of the profiles in which there is no atomic signal. When we subtract the fit curve from the original profiles the resulting profiles have a flat, close-to-zero background. To calculate the atom number at each time-step in the evolution, we integrate the subtracted density profile. For momentum measurements near the beginning and end of the 100-times quench cycle, a fraction of atoms (up to $\sim 10\%$) are lost from the transverse field-of-view as a result of the long TOF used. For these points, we use the number measurement that is closest in time that is not subject to this complication.

{\bf Energy calculations.} To calculate the average energy per atom from a TOF image, we use the normalized density profiles $f(z)$ to weight the energy contributions from each momentum group. The average energy up to a given maximum momentum, $z_m$, is given by 
\begin{equation}\label{energyintegration}
E_{z_m}= \sum_{z_i \in [-z_m, z_m]} {\frac{m}{2}\left(\frac{z_i}{t_\text{TOF}}\right)^{2}f(z_i)\Delta z_i},
\end{equation}
where $t_\text{TOF}$ is the time of flight. Extended Data Fig.~8 shows $E_{z_m}$ for all the points in the first cycle after the 100-times quench. As $z_m$ increases to include all the atoms, the average energy reaches its true value, $E$. We extract $E$ in the face of noise in $f(z)$ by using the average value of $E_{z_m}$ in a region in the flat part of the energy curve. The left cutoff of the region is a 2nd-order polynomial that passes through the origin and the first maxima of the lowest and highest energy curves (the intersections of the polynomial and the energy curves are shown as red circles). The right cutoff is a straight line in the middle of the flat region, which allows us to exclude the high momentum region where there are no atoms, but where the noise is exacerbated by the square in Eq.~(\ref{energyintegration}). We have tried many variants of this $E$ determination, such as using a linear left cutoff, taking left cutoffs at the first maximum of each $E_{z_m}$ curve, and moving around the right cutoff. As long as atoms are not left out and we do not use too high a right cutoff, they give the same energies to within the error we associate with this procedure.

There are three sources of uncertainty in $E$. The first is from the noise in the flat region of $E_{z_m}$ described above. We take the peak-to-peak variation in this region to be the associated error. The other two energy error sources relate to the uncertainty of the background shape in the region where there are atoms. They are most significant near the peak compression points. First, ripples in the background can shift the background level that we infer from our fourth order polynomial fit. To quantify this error, in regions where there are no atoms we subtract the background and filter out noise above $1.25\times 10^{4}$~m$^{-1}$ spatial frequency, which we determine has negligible impact on the background level. We calculate the rms amplitude in the remaining background and then assign an energy error using Eq.~(\ref{energyintegration}), with $f(z_{i})$ replaced by the average rms amplitude. Second, there is uncertainty in the analytical continuation of the 4th-order polynomial into the region with atoms. We analyze the noise spectrum of the background before background subtraction and find the amplitude $A_{t}$ of the characteristic mode specified by the length $L_{t}$ of the region with atoms. The energy error from this source can also be calculated by using Eq.~(\ref{energyintegration}) and replacing $f(z_{i})$ with $f_{t}(z_{i})=A_{t}\cos(\pi z_{i}/L_{t})$. The final energy error bar is the quadrature sum of these three components.

\bibliographystyleM{biblev1}
\bibliographyM{references}

%
%

\vspace{1cm}
\begin{center}
{\bf \large Supplementary Information}
\end{center}

\section{Supplementary Methods}\label{method}

\newpage

\subsection{Energy calculation within GHD}

We define the kinetic energy ($E_K$), the interaction energy ($E_I$), and the total rapidity energy ($E$, which excludes the potential energy in the trap) with respect to the corresponding parts in the Lieb-Liniger Hamiltonian (see Methods). 

We first analyze the homogeneous case, i.e., $U_\ell (z) = 0$, with periodic boundary conditions. In that case, the eigenstates of the Lieb-Liniger Hamiltonian for $N$ bosons are Bethe states~\cite{bethe1931theorie, gaudin2014bethe}, each defined by a different set of rapidities $\{ \theta_1, \dots, \theta_N \}$. (We stress that, in this section, `$\theta_j$' is simply one rapidity in the set defining a Bethe state in a finite-size system; it is not a Fermi point of a multiple Fermi sea.) It is useful to define the density of rapidities associated with a Bethe state,
\begin{equation}
    \label{rapidity_density}
    \rho(\theta) = \frac{1}{L} \sum_{j=1}^N \delta (\theta - \theta_j) ,
\end{equation}
where $L$ is the system size. For finite $N$ and $L$, each eigenstate $\left| \rho \right>$ can be labeled by its rapidity density $\rho(\theta)$. The densities of kinetic and interaction energies in each eigenstate are respectively $e_K = E_K/L =-\frac{\hbar^2}{2m L} \langle\rho| \sum_{j=1}^{N}\partial^2_{z_j} |\rho\rangle$ and $e_I = E_I/L=\frac{g}{L}\langle\rho|\sum_{1\leq j < l \leq N}\delta(z_j-z_l)|\rho\rangle$, and the density of total energy can be written in terms of the rapidity distribution \cite{lieb1963exact}, $e = E/L = \int d\theta \rho(\theta) \frac{\theta^2}{2m}$. The density of interaction energy in the thermodynamic limit $N, L \rightarrow \infty$ can be found in Ref.~\citeSI{kormos2011exact}:
\begin{equation}\label{eI}
e_I= \int d\theta \rho(\theta) \left[\frac{\theta}{m}- v^{\rm eff}(\theta)\right] \theta ,
\end{equation}
where $v^{\rm eff}(\theta)$ is the effective velocity (see Methods). For completeness, we offer a derivation ---different from the one of Ref.~\citeSI{kormos2011exact}--- of this formula below. A direct consequence of Eq.~(\ref{eI}) is that the density of kinetic energy is
\begin{equation}\label{eK}
e_K=e-e_I= \int d\theta \rho(\theta) \left[v^{\rm eff}(\theta)-\frac{\theta}{2m}\right]\theta\,.
\end{equation}

Coming back to the trapped (inhomogeneous) system, we can generalize the previous results within the continuum approximation that underlies the hydrodynamic approach: the system is viewed as a continuum of small fluid cells, each of which is associated with a local distribution $\rho(\theta,z)$. Integrating over the position $z$ along the tubes, and summing the results over all tubes, one obtains the expressions reported in Methods.

{\bf Derivation of Eq.~(\ref{eI}) based on the Hellmann-Feynman theorem.} One can derive Eq.~(\ref{eI}) for the density of interaction energy using the Hellmann-Feynman theorem:
\begin{equation}\label{Hellmann_Feynman}
e_I = \frac{g}{L} \langle\rho| \frac{dH}{dg} |\rho\rangle = g\frac{de}{dg}.
\end{equation}
To calculate $de/dg$, we consider again the homogeneous Lieb-Liniger model [i.e., $U_\ell(z) = 0$] with periodic boundary conditions. We first focus on an eigenstate $\left| \rho \right>$ for finite $N$ and $L$. Each eigenstate depends smoothly on the coupling strength $g$, as the rapidities that define each eigenstate are constrained by the Bethe equations~\cite{gaudin2014bethe, lieb1963exact} which are themselves smooth in $g$ (we set $\hbar = m = 1$, and assume $g>0$),
\begin{equation}\label{bethe}
L\theta_a+\sum_{b=1}^N 2 \, \text{arctan}\left[\frac{\theta_a-\theta_b}{g}\right]=2\pi I_a\, , \qquad a= 1, \dots, N .
\end{equation}
Here the Bethe numbers $I_a$ are integers if $N$ is odd and half-integers if $N$ is even. For each set of non-equal Bethe numbers $\{I_a\}_{1 \leq a \leq N}$, there is a unique solution $\{\theta_a\}_{1 \leq a \leq N}$ to the Bethe equations, which corresponds to one eigenstate of the Lieb-Liniger Hamiltonian. Keeping the set of Bethe numbers fixed, we differentiate Eq.~(\ref{bethe}) with respect to $g$, to get 
\begin{eqnarray}
&&\partial_g\theta_a\bigg[1+\frac{1}{L}\sum_b \varphi(\theta_a-\theta_b)\bigg]\nonumber\\&&=\frac{1}{L}\sum_b\varphi(\theta_a-\theta_b)\bigg(\partial_g\theta_b+\frac{\theta_a-\theta_b}{g}\bigg)\,,
\end{eqnarray}
where $\varphi$ was defined in Methods. We can then take the thermodynamic limit $N, L \rightarrow \infty$, such that the density of rapidities (\ref{rapidity_density}) becomes a continuous function. Using the thermodynamic form of the Bethe equations, $1+\int d\alpha\varphi(\theta-\alpha)\rho(\alpha)=2\pi\rho_s(\theta)$, one finds
\begin{eqnarray}
&& 2\pi\rho_s(\theta_a)\bigg(\partial_g\theta_a-\frac{\theta_a}{g}\bigg) \\ && \nonumber =-\frac{\theta_a}{g}+\int\frac{d\theta}{2\pi}\varphi(\theta_a-\theta_b)n(\theta_b)2\pi\rho_s(\theta_b)\bigg(\partial_g\theta_b-\frac{\theta_b}{g}\bigg)\,. \; \quad
\end{eqnarray}
Rewriting this result using the dressing operation~(see Methods) results in
\begin{equation}
1^{\rm dr} (\theta_a) \left( \partial_g\theta_a-\frac{\theta_a}{g} \right) = - \frac{{\rm id}^{\rm dr} (\theta_a)}{g} ,
\end{equation}
or, equivalently, using the definition of the effective velocity $v^{\rm eff}$ (see Methods),
\begin{equation}
\partial_g\theta_a  =\frac{\theta_a-v^{\rm eff}(\theta_a)}{g}\,.
\end{equation}
Finally, we get the variation of the density of the total energy $e=\frac{1}{L}\sum_a \theta_a^2/2$ in Eq.~(\ref{Hellmann_Feynman}) using the chain rule: $de/dg = \sum_a (\partial_{\theta_a} e ) (\partial_{g} \theta_a)$. This gives Eq.~(\ref{eI}).

\subsection{GHD with a diffusive term}

GHD is a hydrodynamic description of nearly-integrable systems, which is valid in the limit of slow variations and small gradients of rapidity density. Away from that limit, one expects corrections to GHD to become important. In particular, in this work we apply GHD to 1D tubes with small numbers of atoms, such that density variations on lengths of order of the interparticle distance are not small. The most relevant correction is a diffusive term, which turns the GHD equation (see Refs.~\cite{bertini2016transport, castro2016emergent, doyon2019lecture} and Eq.~(3) in Methods) into a Navier-Stokes-like diffusive hydrodynamic equation of the form~\citeM{de2018hydrodynamic, de2019diffusion},~\citeSI{gopalakrishnan2018hydrodynamics,bastianello2020thermalisation}
\begin{equation}
    \label{diffusion}
    \partial_t \rho + \partial_z [v^{\rm eff} \rho] = [\partial_z U(z)] \partial_\theta \rho + \frac{1}{2} \partial_z [\mathfrak{D} \partial_z \rho],
\end{equation}
where the diffusion kernel $\mathfrak{D}$ is a state-dependent linear operator that acts on the rapidity distribution $\rho(\theta)$, and whose exact expression was computed in Ref.~\citeM{de2018hydrodynamic}. In general, Eq.~(\ref{diffusion}) provides a more complete hydrodynamic description of 1D Bose gases than the earlier diffusionless version of Refs.~\cite{bertini2016transport, castro2016emergent}, which is the one used in this work. However, the linear operator $\mathfrak{D}$ of Ref.~\citeM{de2018hydrodynamic} vanishes in zero-entropy states: $\mathfrak{D} = 0$. Therefore, because our simulations involve only zero-entropy states (see Methods), the original diffusionless GHD equation that we use gives the same results as the more complete Eq.~(\ref{diffusion}).

\subsection{Measuring rapidity distributions, and the effect of finite time of flight}

As explained in Methods, in order to experimentally measure the rapidity distribution at time $t$ after the trap quench, we let the atoms expand in a ``flat'' potential in 1D for a time $t_\text{1D}$. We then turn off all confining potentials and let the atoms expand freely in 3D for a time $t_\text{TOF}$. $t_\text{1D}$ and $t_\text{TOF}$ are chosen differently for each trap quench and for different times $t$.

To show the effect of the finite time-of-fight after turning off all confining potentials following the 1D expansions, in Extended Data Fig.~9a we compare the GHD predictions for the total rapidity energy against the ones obtained after finite time-of-fight $t_{\rm TOF}$. To compute the latter, we assume that the local momentum distribution of the gas at $t_\text{1D}$ (whose exact computation requires the use of form factors~\cite{caux2017hydrodynamics} and is beyond the scope of this work) is the rapidity distribution, and we convolve it with a free expansion during a time $t_{\rm TOF}$ to obtain the total rapidity energy that is measured after the time-of-fight $t_{\rm TOF}$. The results in Extended Data Fig.~9a show that the finite $t_{\rm TOF}$ used in our experiments slightly decreases the total rapidity energy in the compression part of the oscillation, while it slightly increases it in the expansion part of the oscillation. In Extended Data Fig.~9b, we show how corrections of this size affect the theory points in Fig.~3b. The shifts are comparable to the experimental error bars on the sides of the curve, and much smaller near the peak. Correcting for this effect does not qualitatively affect the correspondence between the theoretical and experimental points. Since we cannot theoretically calculate momentum distributions, we cannot make similar corrections for the theory in Fig.~3c. We therefore opted not to correct any of the theory points in the main paper for this small effect.

\section{Supplementary Discussion}

\subsection{Validity of GHD for small atom numbers in Tonks-Girardeau limit}

In spite of the excellent agreement between GHD theory and the experimental results for the 100-times quench, one may be wary of the validity of GHD for such small number of atoms per tube (an average of 11) and such a strong quench. Here we check GHD against exact numerical results for a quench in which all parameters are identical to the experimental ones except for $g$, which we take to be $g=7.7\times 10^{-32}$~Jm (i.e., very deep in the Tonks-Girardeau regime, to be compared to $g=3.8\times 10^{-37}$~Jm in the experiments, see Methods). The exact numerical results are obtained by mapping the Tonks-Girardeau bosons onto noninteracting spinless fermions, and solving for the dynamics within the low-density limit in a lattice~\citeSI{xu_rigol_15}, as we did in Ref.~\cite{wilson_malvania_20}  .

Extended Data Fig.~10 shows the exact (solid lines) and GHD results (dashed lines) for $N=5$, 10, and 20 particles, and for the average over all tubes. Except for the fast oscillations, which are Friedel oscillations for the fermionic quasi-particles of the Tonks-Girardeau gas~\citeSI{vignolo2000}, the GHD results for the rapidity distributions in individual tubes match the exact solutions. Also, for the average over tubes, which is what we actually compare to the experiments, the differences between GHD and the exact results are indistinguishable on the scale of the plots. While the only existing way to test GHD with low $N$ and finite coupling strength is by comparison to experiments, as we do, it bolsters confidence in the results that GHD works well with low $N$ and infinite coupling strength.

\bibliographystyleSI{biblev1}
\bibliographySI{references}

%
%

\clearpage

\onecolumngrid

\begin{center}
{\bf \large Extended Data}
\end{center}

\setcounter{figure}{0}

\renewcommand{\figurename}{{\bf Extended Data Figure}}

\begin{figure}[!h]
\includegraphics[width=0.96\columnwidth]{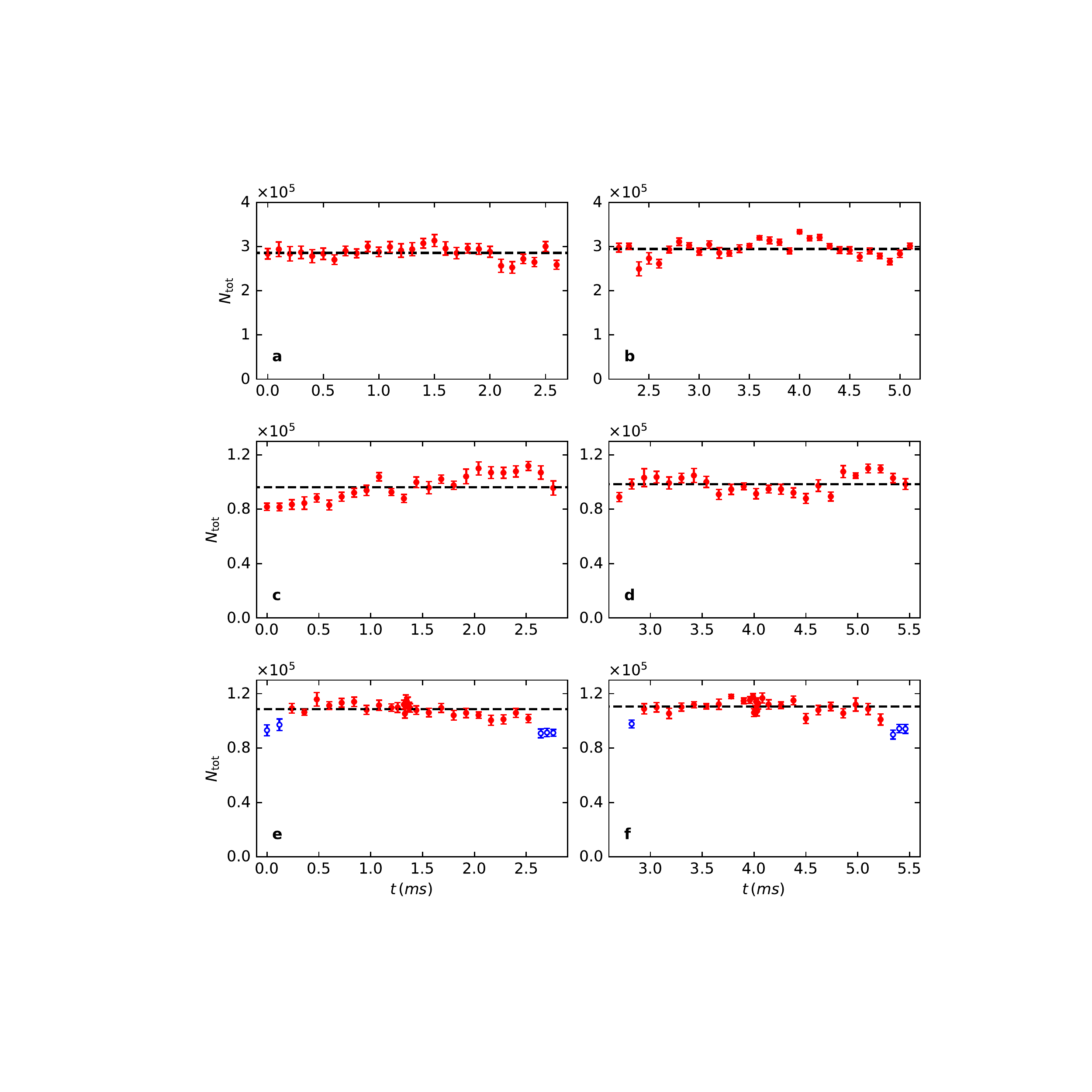}
\caption{{\bf Total atom number extracted from the experimental profiles.} Each point in this figure is, like all the data points in the paper, an average over 15 runs of the experiment and the error bars are the standard deviation. textbf{a} and \textbf{b}, First and second cycle, respectively, of the 10-times quench rapidity measurement. \textbf{c} and \textbf{d}, First and second cycle, respectively, of the 100-times quench rapidity measurement. \textbf{e} and \textbf{f}, First and second cycle, respectively, of the 100-times quench momentum measurement. The empty (blue) symbols in \textbf{e} and \textbf{f} indicate data points for which up to 10\% of the atoms leave the imaging field-of-view, thus $N_{\rm tot}$ does not reflect the total number of atoms. For these points, we use the number measurement that is closest in time that is not subject to this complication. Black dashed lines in each panel show the averaged atom number [not including the empty (blue) symbols]. For the GHD simulations described in the main text, we use $N_{\rm tot}$ in a range from 250,000 (83,000) to 330,000 (119,000) in steps of 10,000 (3,000) atoms for the 10-(100-)times quench. The measured atom number for each individual evolution time is rounded to the simulation with closest $N_{\rm tot}$. The step sizes are about as big as the error bars of the number measurements.}
\label{totalatom}
\end{figure}

\begin{figure}[!t]
\includegraphics[width=0.99\columnwidth]{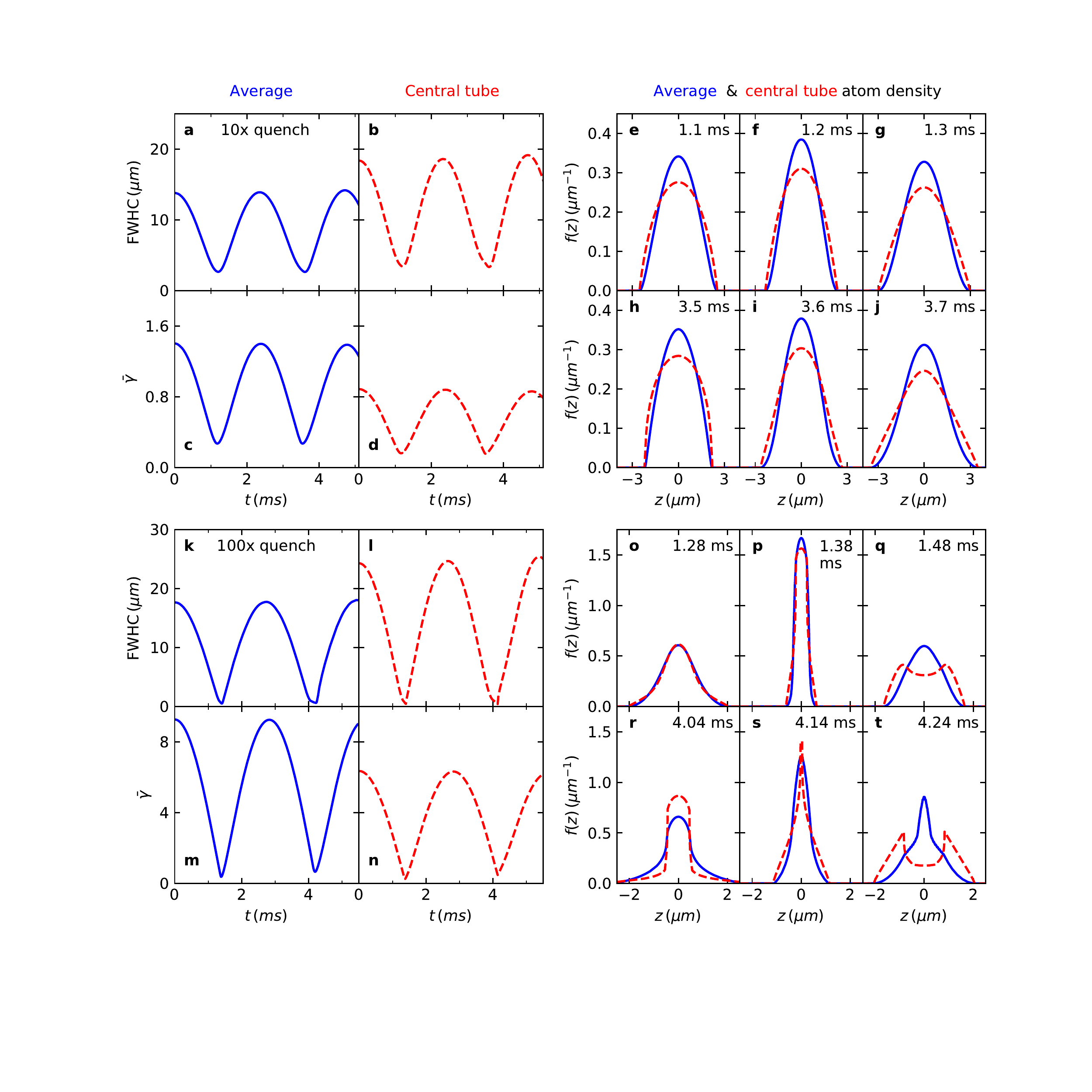}
\caption{{\bf Evolution of the atom density and related quantities.} {\bf a}--{\bf j} ({\bf k}--{\bf t}), Theoretical results for the 10-(100-)times quench with $N_{\rm tot}=300,000$ (98,000) using the same trap parameters as for the first period in Fig.~2b (Fig.~3b). \textbf{a} and \textbf{k} show the evolution of the full width at half the central atom density (FWHC) for the average over all tubes. \textbf{b} and \textbf{l} show the same for the central tube in each case (which have 140 and 24 atoms, respectively). \textbf{c} and \textbf{m} show how $\bar\gamma$ evolves when averaged over all tubes. \textbf{d} and \textbf{n} show how $\bar\gamma$ evolves in the central tube. The continuous (dashed) lines in \textbf{e}--\textbf{j} and in \textbf{o}--\textbf{q} show the normalized atom density $f(x)$ for the average over all tubes (central tube) at times near the maximum compression points in the first (\textbf{e}--\textbf{g} and \textbf{o}--\textbf{q}) and second (\textbf{h}--\textbf{j} and \textbf{r}--\textbf{t}) cycles. The times shown for the 10-times quench are $t=$1.1, 1.2, 1.3 ms (\textbf{e}--\textbf{g}) and $t=$3.5, 3.6, 3.7 ms (\textbf{h}--\textbf{j}). For the 100-times quench they are $t=$1.28, 1.38, 1.48 ms (\textbf{o}--\textbf{q}) and $t=$4.04, 4.14, 4.24 ms (\textbf{r}--\textbf{t}), which are the same times for which the Fermi occupation ratios for the central tube are shown in Extended Data Fig.~6. The presence of multiple peaks in the atom densities (as in, e.g., \textbf{q} and \textbf{t}) dictates the use of the FWHC in panels \textbf{a}, \textbf{b}, \textbf{k} and \textbf{l}, as opposed to the more common full width at half maximum.}
\label{gamma_FWHM}
\end{figure}

\begin{figure}[!t]
\includegraphics[width=0.99\columnwidth]{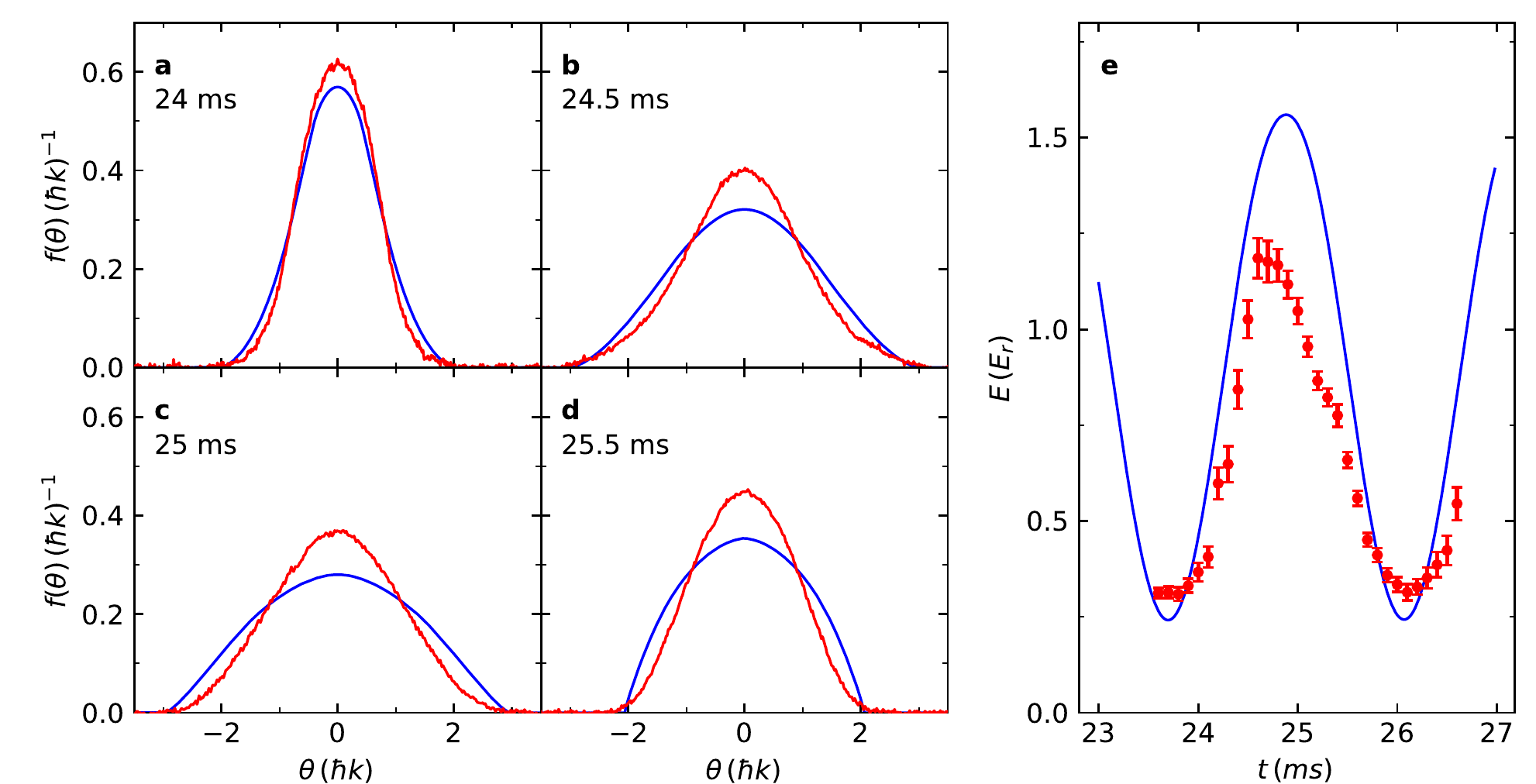}
\caption{{\bf The 11th oscillation cycle after the 10-times quench.} {\bf a}-{\bf d}, Experimental (red lines) and GHD theory (blue lines) rapidity distributions at the corresponding times. {\bf e}, Rapidity energies for the experiment (red symbols) and GHD theory (blue lines). The total atom number used for the theory is 300,000, the average initial atom number. The average atom number measured in the 11th oscillation cycle is 240,000. The theory does not account for the atom loss we observe during the time evolution. The trap depth is adjusted to match the phase of oscillation. The way the energy is reduced in the experiment can be qualitatively understood because the inelastic loss from 3-body collisions disproportionately occurs at the peak compression points. That is therefore when the interaction energy is lost. There is then less interaction energy to drive the expansions, so the resulting smaller cloud length offsets the lower atom number at the peak expansion points. Therefore the energy minima change little from cycle to cycle, while the energy maxima decrease.}
\label{9x11cycle}
\end{figure}

\begin{figure}[!t]
\includegraphics[width=0.99\columnwidth]{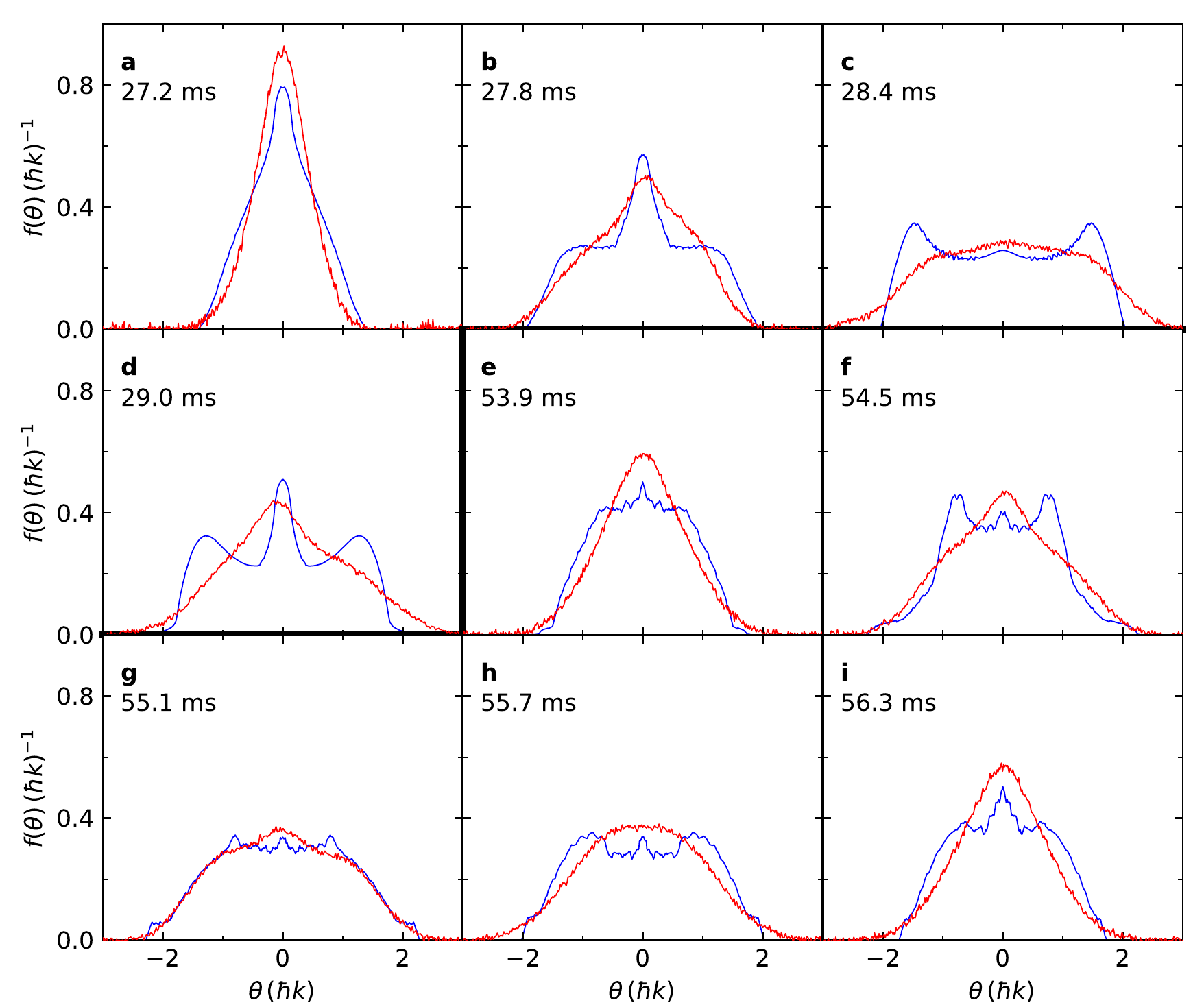}
\caption{{\bf Rapidity profiles during the 11th (\textbf{a}-\textbf{d}) and 21st (\textbf{e}-\textbf{i}) cycles of the 100-times quench.} The red curves are the experimental data and the blue curves are GHD simulations with $N_{\rm tot}=113,000$. The trap depth is adjusted (once for each cycle) so that the theoretical and experimental oscillations are in phase. It is important to stress that our theoretical modeling is done directly with GHD; it is not based on a full quantum simulation of the Lieb-Liniger model, which is not currently possible. GHD is a hydrodynamic theory that only provides an approximation to the predictions of the Lieb-Liniger model. As time increases, discrepancies between GHD and full quantum simulations of the Lieb-Liniger model are expected to increase. Similarly, small experimental imperfections may wash out sharper features as time increases. There is no clear way to estimate the extent of these accumulated errors, and so it is not possible to identify how much of the disagreement comes from the theoretical approximations and how much comes from experimental imperfections.}
\label{100xlaterprofiles}
\end{figure}

\begin{figure}[!t]
\includegraphics[width=1\columnwidth]{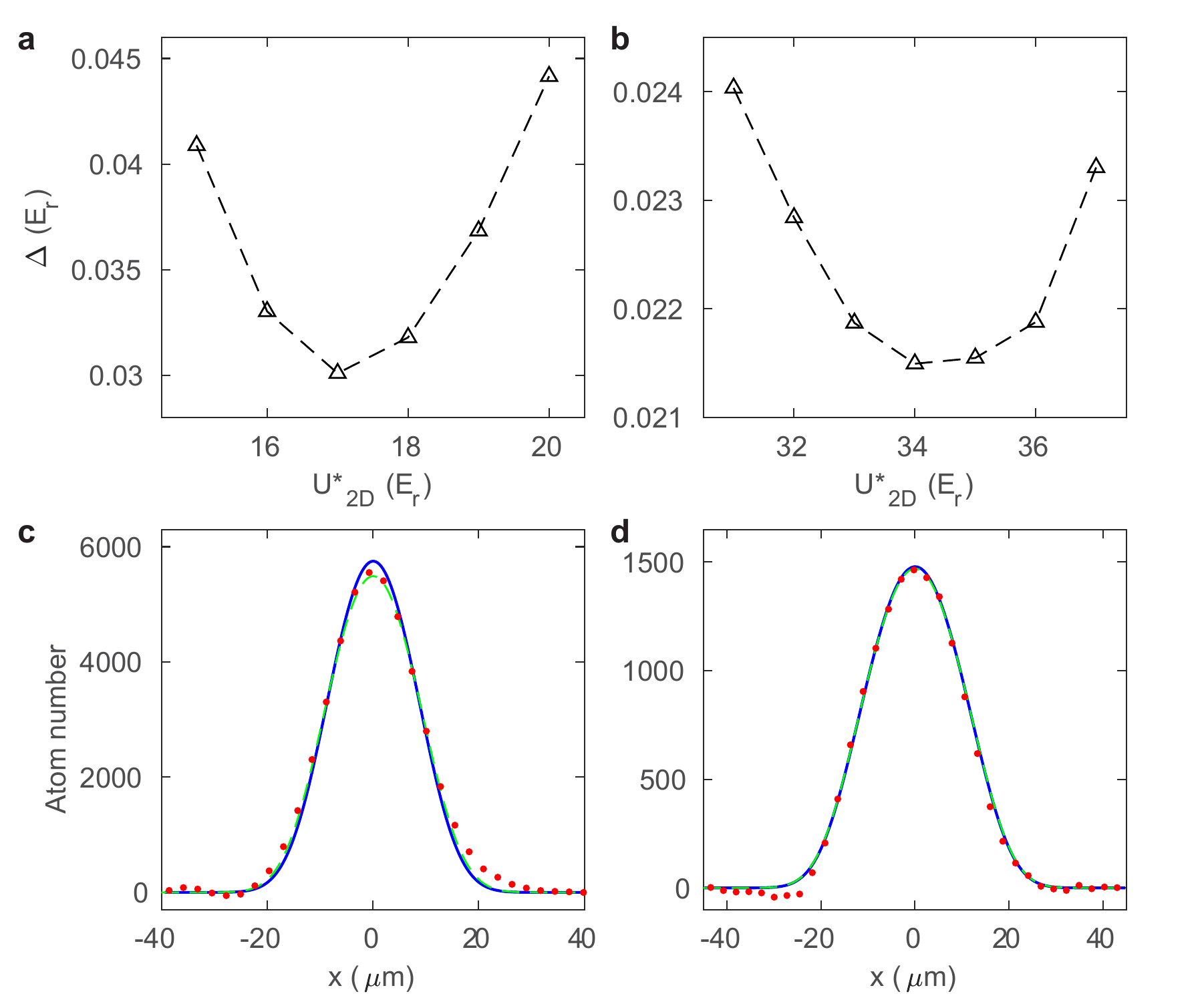}
\caption{{\bf Determination of the transverse distribution of 1D gases.} \textbf{a},  The root mean square (RMS) difference between the measured rapidity energy and the GHD simulation for the intermediate coupling initial condition. The difference is plotted as a function of the 2D lattice depth at which the 1D gases decouple, $U^*_{\rm 2D}$. Because they are more sensitive to small density changes, only the seven data points around the first rapidity energy peak are used for the RMS calculation. \textbf{b}, Same as \textbf{a}, but for the strong coupling initial condition and with six data points for the RMS calculation. \textbf{c}, Comparison of the measured initial transverse atom distribution for intermediate coupling (red points) to the distribution determined from the measurement described in \textbf{a} (blue line), convolved with the 4.8 $\mu$m instrumental resolution.  We also show the 1D projection of a 3D Thomas-Fermi distribution (green dashed line), where the Thomas-Fermi size of 17 $\mu$m is obtained by fitting the distribution, convolved with the instrumental resolution, to experimental data between -13 $\mu$m and 13 $\mu$m. \textbf{d}, Same as \textbf{c} but for the strong coupling initial condition. For the blue line, $U^*_{\rm 2D}=34E_r$, as determined from \textbf{b}. For the green dashed curve the best-fit Thomas-Fermi size is 22 $\mu$m. In \textbf{c} and \textbf{d}, one can see in the tails of the experimental distributions the effect of slight lensing in the absorption imaging measurement. (This is only present at the high densities of an in situ measurement, not in any of our momentum or rapidity measurements). The blue curves take advantage of the high sensitivity of the rapidity measurement on density, and they give a result that is consistent with the direct density measurement. The green curves are included here because we have previously used that method to determine transverse distributions (e.g., in Ref.~[4] in the main text). The new method is more sensitive and more theoretically consistent with what happens as the 2D lattice is turned on. Still, \textbf{c} and especially \textbf{d} show that the difference in the two methods ends up being quite modest.}
\label{Atomdistribution}
\end{figure}

\begin{figure}[!t]
\includegraphics[width=1\columnwidth]{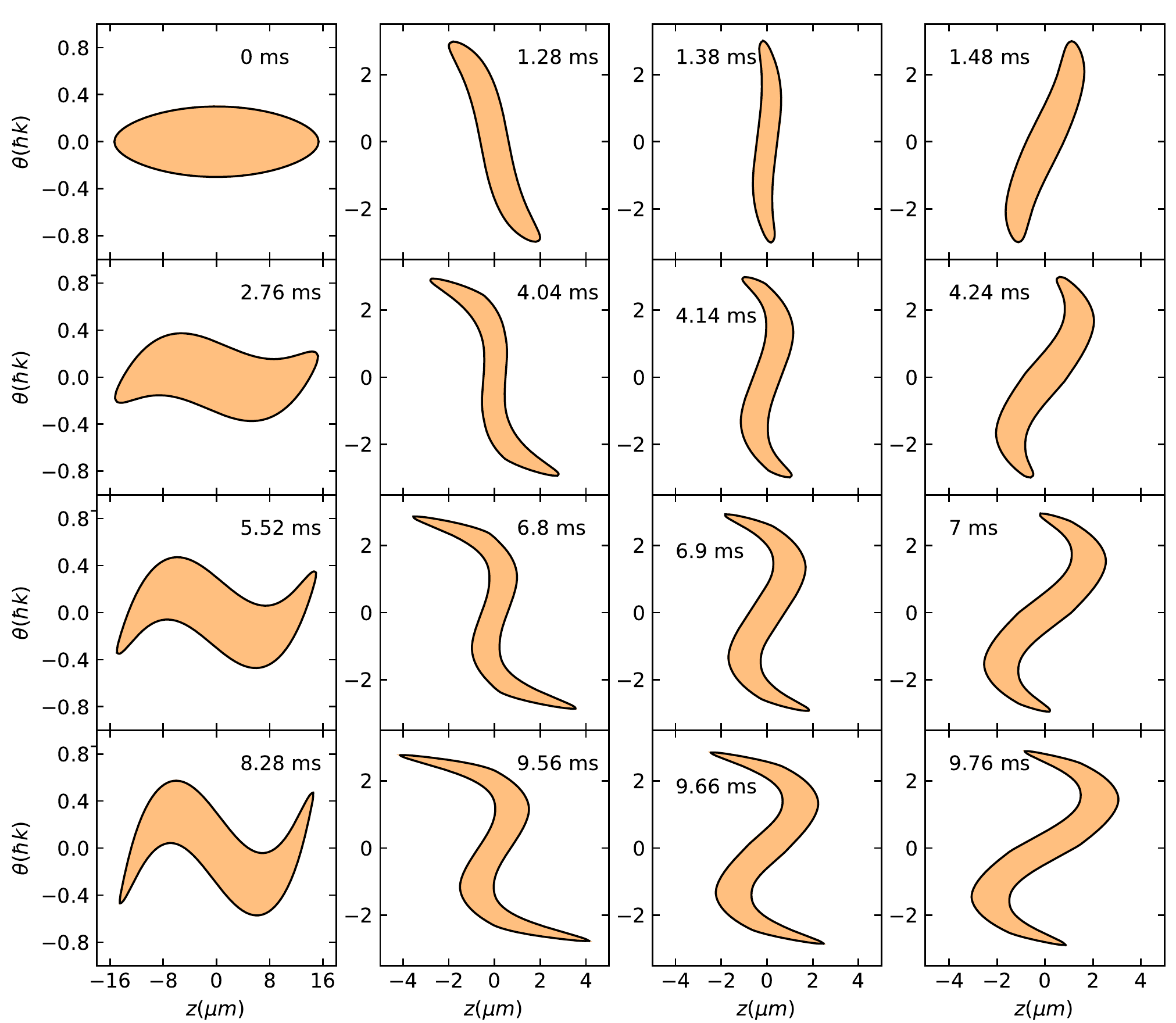}
\caption{{\bf Evolution of the Fermi occupation ratio $n(\theta,z)$ after the 100-times trap quench.} We show the GHD simulation over the first four breathing cycles for a single tube with atom number $N=24$, initial trap depth $U_0=1.195\,E_r$, final trap depth $U_f=125.7\,E_r$, and Gaussian trap width $W=55.2\,\mu$m. The occupation ratio is $n(\theta,z) = 1$ inside the colored area, and $0$ outside. The boundary of the area where $n(\theta,z) = 1$ is a closed curve $\Gamma_t$, whose evolution is computed using the GHD equations (see Methods). For clarity, we adjust the range in $z$ and $\theta$. When a vertical line can be drawn that passes through the boundary more than twice, there is more than one Fermi sea and GHD is necessary to solve for the dynamics. For instance, a second Fermi sea forms first near the first compression point (1.38~ms), becoming more distinct at later peak compression points. A third Fermi sea first forms in the vicinity of 4.2~ms. From among these snapshots it can first be seen (barely) in the 6.9 ms panel.}
\label{fermisurface}
\end{figure}

\begin{figure}[h]
\includegraphics[width=0.65\columnwidth]{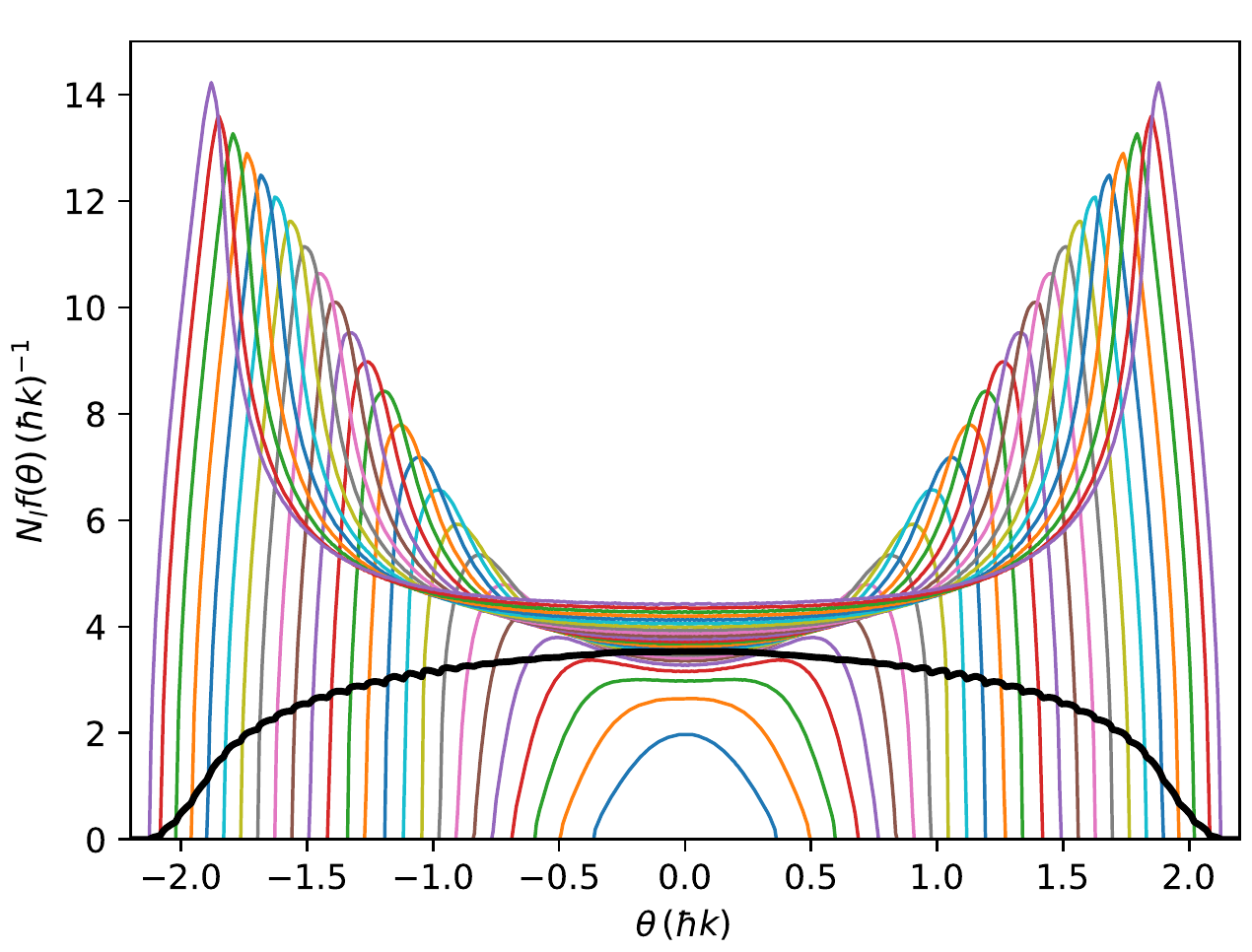}
\caption{{\bf Rapidity distributions for 14.4~ms after the 100-times quench.} The time corresponds to a point in the 6th breathing cycle (see Fig.~4c in the main text). The rapidity distributions are plotted for individual tubes with atom number $N_\ell=1$, 2, \ldots, 25 (colored lines from bottom to top). The thick black line is the average over all the tubes. See the discussion following Eq.~(9) in Methods.}
\label{ripples}
\end{figure}

\begin{figure}[!t]
\includegraphics[width=1\columnwidth]{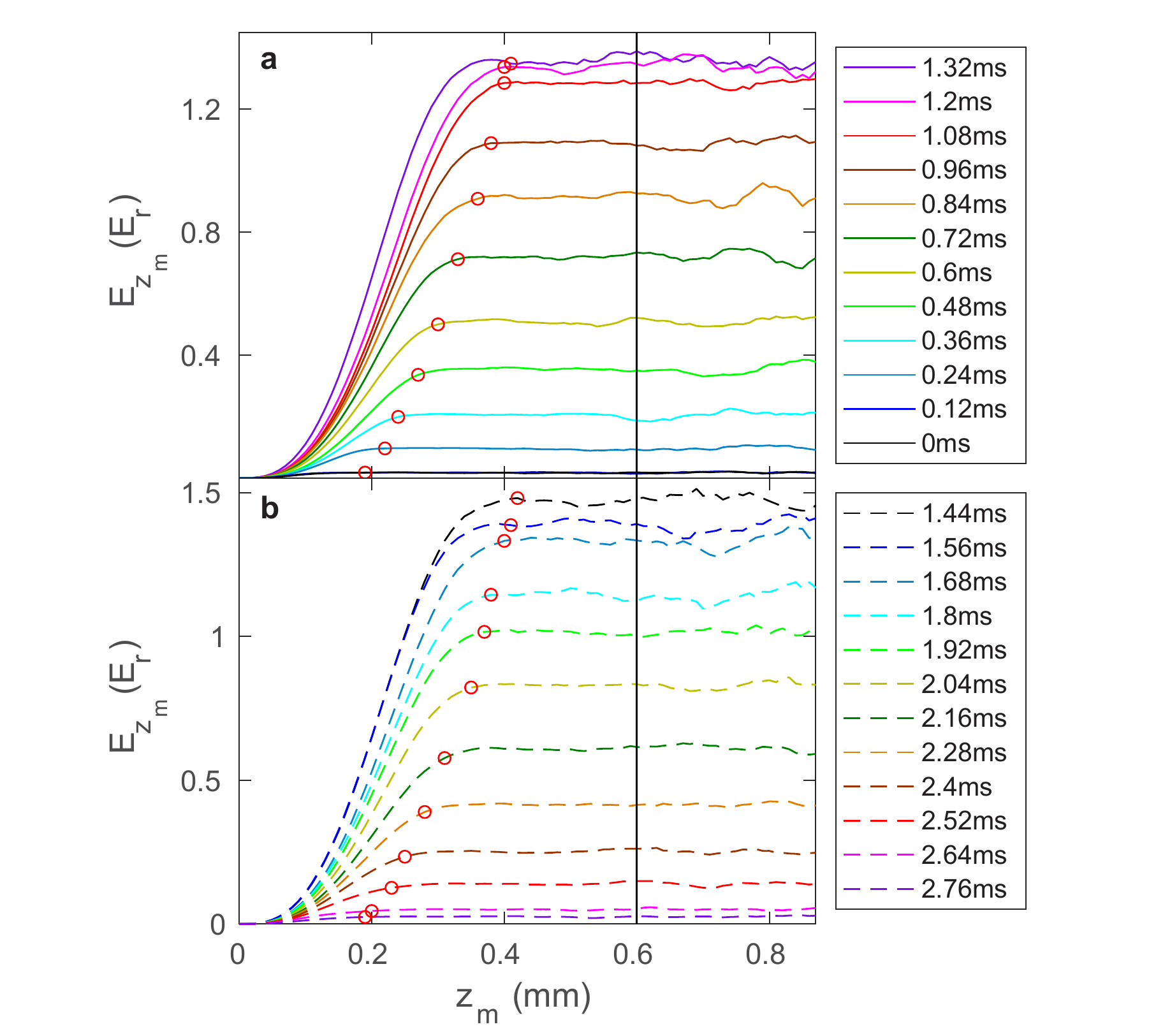}
\caption{\textbf{Average rapidities energy vs.~the integration limit.} \textbf{a}, Data from the first half cycle after the 100-times quench. \textbf{b}, Data from the second half cycle after the 100-times quench. The energy curve (solid lines in \textbf{a}, dashed lines in \textbf{b}) at each evolution time is obtained by integrating symmetrically about the center of the corresponding profile up to $z_m$. Every energy curve eventually flattens out when all the atoms have been included. For each curve, $E$ is taken to be the average value of $E_z$ between the left cutoff (red circle) an the right cutoff (black line); the procedure for determining the cutoffs is described in Methods. The part of the uncertainty in $E$ due to this procedure is given by the range of $E_z$ between the cutoffs.}
\label{energy_cutoff}
\end{figure}

\begin{figure}[!t]
\includegraphics[width=0.7\columnwidth]{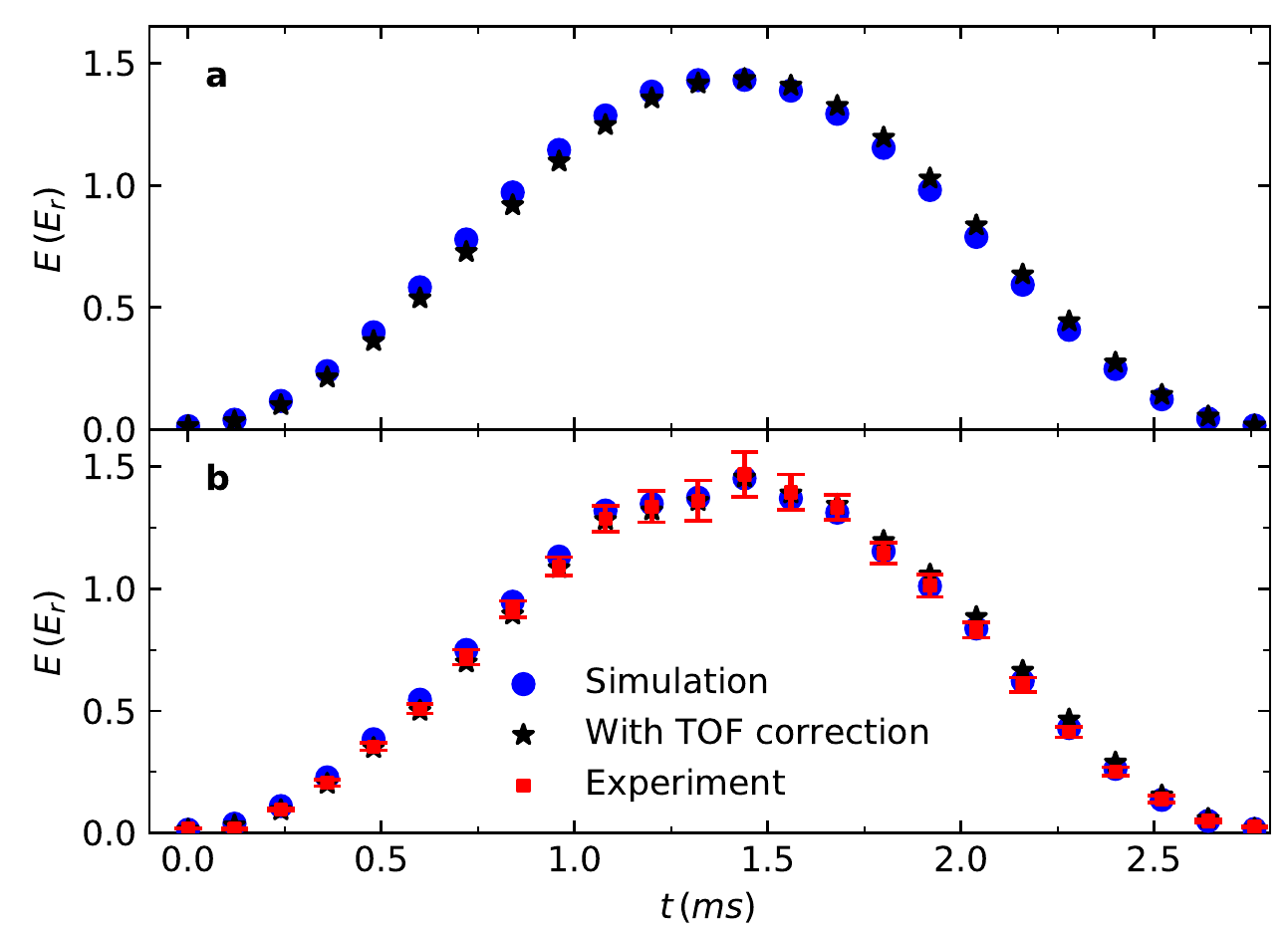}
\caption{\textbf{Expected effect of the finite TOF expansion for the 100-times trap quench.} {\bf a}, GHD simulations for a system with $N_{\rm tot}=98,000$ and the same trap parameters as for the first period in Fig.~3b in the main text. The blue circles show the rapidity energy and the black stars show the inferred rapidity energy after TOF (see Supplementary information). Finite TOF leads to a small energy reduction before the peak compression and a small energy increase after. \textbf{b}, GHD results accounting for the measured atom number (blue circles, also shown in Fig.~3b in the main text), the GHD results corrected by the TOF effect from {\bf a} (black stars), and the experimental results (red squares, also shown in Fig.~3b in the main text). The finite TOF effect is at most comparable to the error bars, so rather than making what would have to be a small ad hoc correction to the theoretical points in the rapidity and kinetic energies in the main text, we neglect it.}
\label{TOF100x}
\end{figure}

\begin{figure}[h]
\includegraphics[width=0.99\columnwidth]{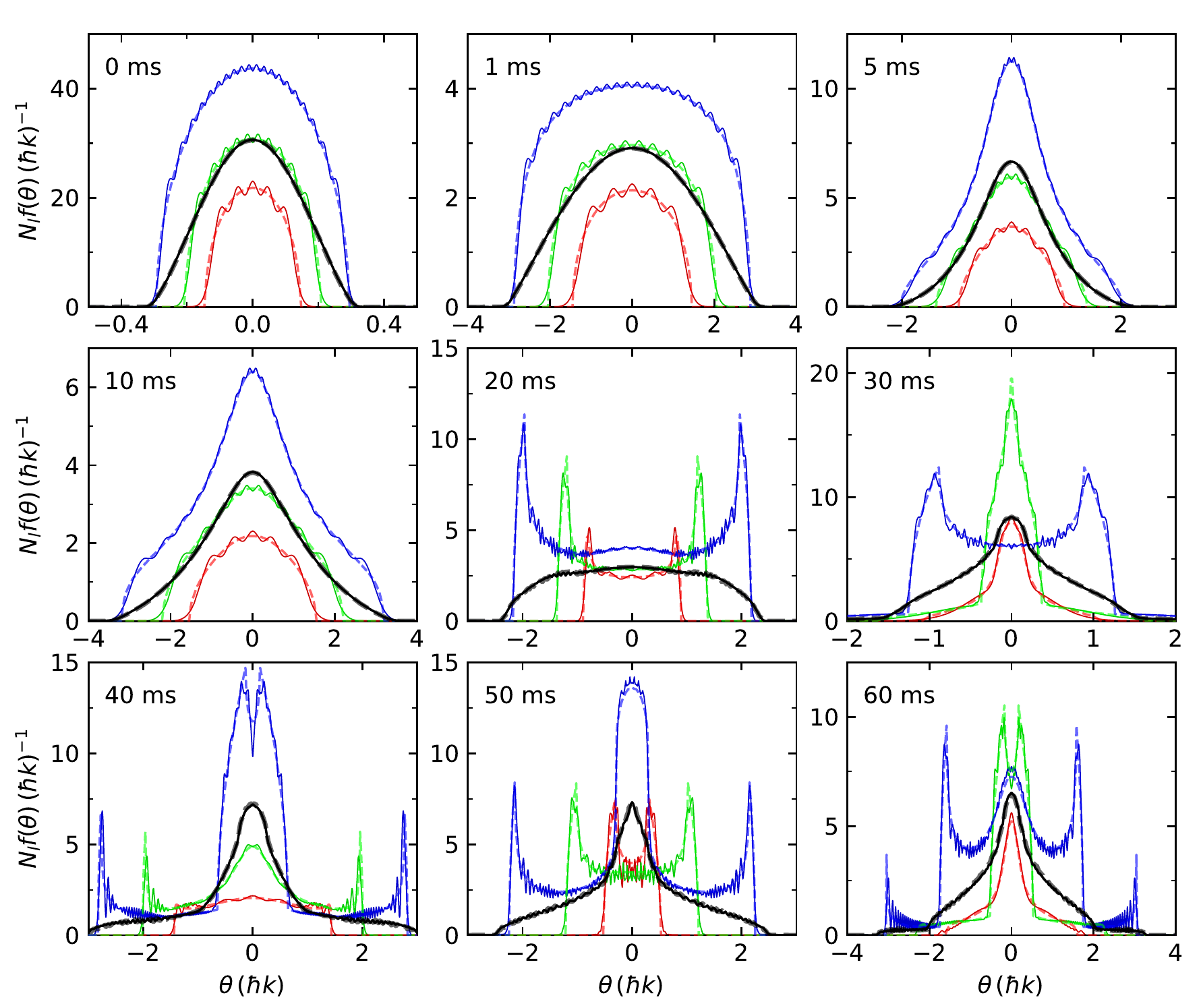}
\caption{\textbf{Comparison of rapidity profiles from GHD and from exact dynamics in the Tonks-Girardeau limit, for small atom numbers.} All parameters are identical to the experimental ones for the first cycle of the 100-times quench (main text Fig.~3b) except for $g$, which we take to be $g=7.7\times 10^{-32}$~Jm (i.e., very deep in the Tonks-Girardeau regime, to be compared to $g=3.8\times 10^{-37}$~Jm in the experiments, see Methods). In each panel, the solid lines are for the exact dynamics and the dashed lines are the GHD results. The red, green, and blue lines are for single tubes with 5, 10, and 20 atoms, respectively. The black lines are obtained by averaging the results over all tubes in the 2D array. Apart from the ripples associated with atom quantization, the GHD results overlap the exact ones, even for as few as 5 particles.}
\label{TG_GHD_Exact}
\end{figure}

\end{document}